\documentclass[prd,amsmath,aps,floats,amssymb, floatfix, superscriptaddress,nofootinbib,twocolumn,preprintnumbers]{revtex4-1}  

\setcitestyle{authoryear,round}
\usepackage[T1]{fontenc}
\usepackage{aecompl}

\usepackage{newtxtext,newtxmath}
\usepackage{mdframed,bm}
\usepackage[T1]{fontenc}
\usepackage{ae,aecompl}
\usepackage{graphicx}	
\usepackage{amsmath}	
\usepackage{amssymb}	
\usepackage{booktabs}
\usepackage{color}
\usepackage{enumitem}
\usepackage[colorlinks=true]{hyperref}

\hypersetup{
    urlcolor=black,
    citecolor=blue,
    linkcolor=blue,
  }%

\usepackage{physics}
\usepackage{cleveref}
\usepackage{comment}
\usepackage[dvipsnames]{xcolor}

\usepackage{xspace}
\usepackage{xcolor}
\usepackage{float}
\restylefloat{table}

\usepackage{physics}

\newcommand{\apjs}{ApJS}

\newcommand{\mnras}{MNRAS}

\newcommand{\pasp}{PASP}

\DeclareRobustCommand{\PRF}[3]{#2}

\DeclareRobustCommand{\PRF}[3]{#2} 
\definecolor{bostonuniversityred}{rgb}{0.8, 0.0, 0.0}

\allowdisplaybreaks
\begin{document}

\title{Dark Energy Survey Year 3 Results: \\ Exploiting small-scale information with lensing shear ratios}

\author{C.~S{\'a}nchez} \email{carles.sanchez.alonso@gmail.com}
\affiliation{Department of Physics and Astronomy, University of Pennsylvania, Philadelphia, PA 19104, USA}
\author{J.~Prat} \email{jpratmarti@gmail.com}
\affiliation{Department of Astronomy and Astrophysics, University of Chicago, Chicago, IL 60637, USA}
\affiliation{Kavli Institute for Cosmological Physics, University of Chicago, Chicago, IL 60637, USA}
\author{G.~Zacharegkas}
\affiliation{Kavli Institute for Cosmological Physics, University of Chicago, Chicago, IL 60637, USA}
\author{S.~Pandey}
\affiliation{Department of Physics and Astronomy, University of Pennsylvania, Philadelphia, PA 19104, USA}
\author{E.~Baxter}
\affiliation{Institute for Astronomy, University of Hawai'i, 2680 Woodlawn Drive, Honolulu, HI 96822, USA}
\author{G.~M.~Bernstein}
\affiliation{Department of Physics and Astronomy, University of Pennsylvania, Philadelphia, PA 19104, USA}
\author{J.~Blazek}
\affiliation{Department of Physics, Northeastern University, Boston, MA 02115, USA}
\affiliation{Laboratory of Astrophysics, \'Ecole Polytechnique F\'ed\'erale de Lausanne (EPFL), Observatoire de Sauverny, 1290 Versoix, Switzerland}
\author{R.~Cawthon}
\affiliation{Physics Department, 2320 Chamberlin Hall, University of Wisconsin-Madison, 1150 University Avenue Madison, WI  53706-1390}
\author{C.~Chang}
\affiliation{Department of Astronomy and Astrophysics, University of Chicago, Chicago, IL 60637, USA}
\affiliation{Kavli Institute for Cosmological Physics, University of Chicago, Chicago, IL 60637, USA}
\author{E.~Krause}
\affiliation{Department of Astronomy/Steward Observatory, University of Arizona, 933 North Cherry Avenue, Tucson, AZ 85721-0065, USA}
% ...
\collaboration{DES Collaboration}

\date{\today}

\vspace{10mm}

\label{firstpage}
\begin{abstract}
Using the first three years of data from the Dark Energy Survey (DES), we use ratios of small-scale galaxy-galaxy lensing measurements around the same lens sample to constrain source redshift uncertainties, intrinsic alignments and other systematics or nuisance parameters of our model. Instead of using a simple geometric approach for the ratios as has been done in the past, we use the full modeling of the galaxy-galaxy lensing measurements, including the corresponding integration over the power spectrum and the contributions from intrinsic alignments and lens magnification. We perform extensive testing of the small-scale shear ratio (SR) modeling by studying the impact of different effects such as the inclusion of baryonic physics, non-linear biasing, halo occupation distribution (HOD) descriptions and lens magnification, among others, and using realistic $N$-body simulations of the DES data. We validate the robustness of our constraints in the data by using two independent lens samples with different galaxy properties, and by deriving constraints using the corresponding large-scale ratios for which the modeling is simpler. The results applied to the DES Y3 data demonstrate how the ratios provide significant improvements in constraining power for several nuisance parameters in our model, especially on source redshift calibration and intrinsic alignments (IA). For source redshifts, SR improves the constraints from the prior by up to 38\% in some redshift bins. Such improvements, and especially the constraints it provides on IA, translate to tighter cosmological constraints when shear ratios are combined with cosmic shear and other 2pt functions. In particular, for the DES Y3 data, SR improves $S_8$ constraints from cosmic shear by up to 31\%, and for the full combination of probes (3$\times$2pt) by up to 10\%. The shear ratios presented in this work are used as an additional likelihood for cosmic shear, 2$\times$2pt and the full 3$\times$2pt in the fiducial DES Y3 cosmological analysis. \\
\vspace{25pt}

\end{abstract}

\preprint{DES-2020-0596}
\preprint{FERMILAB-PUB-21-247-AE}
\maketitle



\section{Introduction}

As photons from a distant light source travel through the Universe, their paths are perturbed by the gravitational influence of the large-scale structure. Weak gravitational lensing concerns the small distortions in the images of distant galaxies due to the influence of the intervening mass along the line of sight (see e.g.~\citealt{Kilbinger2009} for a review). In particular, galaxy-galaxy lensing (or simply galaxy-shear) refers to the correlation between foreground (lens) galaxy positions and the tangential component of lensing shear of background (source) galaxies at higher redshifts, which is a measure of the projected, excess mass distribution around the lens galaxies \citep{Bardeen1986}. Extracting useful cosmological or astrophysical information from galaxy-galaxy lensing is complicated by a number of factors. First, one needs to model the relationship between the galaxy density field and the underlying matter field, i.e. galaxy bias \citep{Fry1993}. Second, at small angular separations between lens and source, the signal-to-noise tends to be large, but lensing-galaxy two-point functions become increasingly sensitive to the small-scale matter power spectrum, whose modeling is convoluted due to non-linearities and baryonic effects \citep{van2014impact,semboloni2013effect,harnois2015}. Also, galaxy bias may become scale-dependent at those scales (e.g. \citealt{Cresswell_2008}). To sidestep these limitations, several studies in the past have considered the usage of ratios between galaxy-shear two-point functions sharing the same lens sample, also called lensing ratios. This observable cancels out the dependence on the galaxy-matter power spectrum while keeping the sensitivity to the angular diameter distances of both tracer and source galaxies.

Several applications of lensing ratios have been considered in the literature. They were originally proposed in \citet{Jain2003} as a novel way to constrain cosmology from geometrical information only, using ratios of galaxy-shear cross-correlation functions sharing the same lens sample. They envisioned dark energy properties could be constrained using these ratios, in particular the parameter describing the equation of state of dark energy, $w$. \citet{Taylor_2007} proposed applying this technique behind clusters using ratios of individual shear measurements, rather than correlation functions. This revised method was applied to data in \citet{Kitching_2007} using lensing measurements around three galaxy clusters, obtaining weak constraints on $w$. Later, \citet{Taylor_2012} used low-mass systems from the HST Cosmos Survey and was able to detect cosmic acceleration.  Other authors developed variants of these initial methods, including \citet{Zhang_2005}, who proposed an approach for both galaxy-shear and shear-shear correlations. 
Also, \citet{Bernstein_2004} explored an alternative formalism for implementing the original idea of \citet{Jain2003}, and documented for the first time that the dependence on cosmology was rather weak. They showed that to achieve sensitivity on cosmological parameters, photometric redshifts had to be extremely well characterized, together with the calibration of shear biases, unless they were redshift independent. \citet{Kitching_2008} also discussed systematics affecting shear-ratio in detail, also finding that photometric redshift uncertainties played a prominent role.

Given this dominant dependency on photometric redshift uncertainties, lensing ratios of galaxy-galaxy lensing measurements have been established as a probe to test redshift distributions and redshift-dependent multiplicative biases \citep{Schneider_2016}. Note that in combination with CMB lensing, geometrical lensing ratios \textit{can} still constrain cosmological parameters (\citealt{Das_Spergel}, \citealt{kitching2015rcslens}, \citealt{Singh2016}, \citealt{Miyatake2017}, \citealt*{Prat_2019} ), but otherwise they have been found to be dominated by redshift uncertainties. Because of that, many studies have used shear ratios to cross-check the redshift distributions of the source sample computed with another method. This is what is known as the ``shear-ratio test'', where ratios of galaxy-galaxy lensing measurements are used to test the redshift distributions of different redshift bins for the corresponding shape catalog. This has been done in several galaxy surveys such as in the Sloan Digital Sky Survey (SDSS), e.g. \citet{Mandelbaum_2005}, where both redshifts and multiplicative shear biases were tested for the first time, in the Red-Sequence Cluster Survey (RCS) \citep{Hoekstra_2005} and in the Kilo-Degree Survey (KiDS)  \citep{Heymans_2012,Hildebrandt_2017,Hildebrandt_2020,Giblin_2020}. 

In the Dark Energy Survey (DES) Y1 galaxy-galaxy lensing analysis \citep*{Y1GGL}, geometrical lensing ratios were used to place constraints on the redshift distributions of the source samples and obtained competitive constraints on the mean of the source redshift distributions. This was among the first times the shear-ratio information was used to place constraints instead of just as a diagnostic test. They were also able to constrain multiplicative shear biases. In the current study, we continue this line of work, but generalize this approach in several ways. We develop a novel method that uses lensing ratios as an extra probe to the combination of galaxy-galaxy lensing, cosmic shear and galaxy clustering, usually referred to as 3$\times$2pt. Specifically, we add the \textit{shear-ratio likelihood}, that uses small-scale independent information, to the usual 3$\times$2pt likelihood. This extra likelihood places constraints on a number of astrophysical parameters, not only those characterizing redshift uncertainties but, importantly, also those characterizing intrinsic alignments and multiplicative shear biases at the same time. By helping to constrain these nuisance parameters, the lensing ratios at small scales provide additional information to obtain tighter cosmological constraints, while still being insensitive to baryonic effects and non-linear galaxy bias. 

Using the first three years of observations from DES (Y3 data), we construct a set of ratios of tangential shear measurements of different source redshift bins sharing the same lens bin, for different lens redshift bins. These ratios have the advantage that they can be modeled in the small, non-linear scale regime where we are not able to accurately model the original two-point correlation functions and which is usually discarded in cosmological analyses. This allows us to exploit information from small scales which would have otherwise been disregarded given our inability to model the tangential shear at small scales due to uncertainties in the galaxy bias model, the matter power spectrum, baryonic effects, etc, which cancel out in the ratios. This cancellation happens exactly only in the limit where the lens redshift distribution is infinitely narrow which is when lensing ratios can be perfectly modeled with geometry only. Instead, if the lens redshift distribution has some finite width, as happens in realistic scenarios as in this work, the cancellation is not exact and the ratios retain some dependence on the lens properties and matter power spectrum, though still much smaller than in the tangential shear signal itself. There are further effects which introduce dependence of shear ratios on parameters other than cosmological distances, such as magnification of the lens galaxies, and the alignment of the source galaxy orientations with the lens galaxy positions due to their physical association, what is usually referred to as Intrinsic Alignments (IA).

There are several different approaches to account for magnification and IA effects on shear ratios. One possible approach is to mitigate these effects in the ratios, e.g.~recently \citet{Unruh_2019} proposed a mitigation strategy for lens magnification effects in the shear-ratio test. Another option is to include these effects in the model, e.g. \citet{Giblin_2020} performed a shear-ratio test on the latest KiDS data set and included non-linear alignment (NLA, \citealt{Hirata2004,Bridle2007}) intrinsic alignment terms in their originally geometrical model. Importantly, they note that SR is indeed very sensitive to the IA model, and they suggest the combination of SR with other cosmological observables to fully exploit the IA constraining power of SR. 

In this work we develop a SR analysis that takes full advantage of the IA dependence of the probe, and combine it with other observables to fully exploit the gains in cosmological constraining power. We can do that by describing the ratios using the full tangential shear model, as it is used in the DES Y3 3$\times$2pt analysis, but on smaller scales. In this way, we do not only take into account the width of the lens redshift distributions but also lens magnification and intrinsic alignment effects. Moreover, this original approach also has the advantage of not adding extra computational cost: the $3\times2$pt analysis already requires calculation of the full galaxy-galaxy lensing model for all the scales and source/lens combinations that we use.

Thus, the approach we develop in this work can be thought of extending the galaxy-galaxy lensing data-vector to smaller scales, where most of the signal-to-noise lies, but using the ratio transformation to retain the information we can confidently model. The threshold scale where we are not able to model DES Y3 tangential shear measurements accurately enough given the current uncertainties has been set at 6 Mpc/$h$ \citep{Krause_2017} for the 3$\times$2pt analysis. The ratios we use in this work only use tangential shear measurements below this threshold, to provide independent information. The small-scale limit of the ratios is set by the regime of validation of our IA model, in some cases, and otherwise by the angular range that has been validated for galaxy-galaxy lensing \citep{y3-gglensing}.

In this paper we explore the constraining power of lensing ratios first by themselves and then in combination with other probes such as galaxy clustering, galaxy-galaxy lensing and cosmic shear. We use the same model setup as in the DES Y3 3$\times$2pt cosmological analysis \citep{y3-3x2ptkp}, using the same nuisance parameters including IA, lens and source redshift parameters and multiplicative shear biases. We test this configuration first using simulated data vectors and $N$-body simulations to then apply it to DES Y3 data. We perform a series of tests to validate our fiducial model against different effects which are not included in it such as the impact of baryons, non-linear bias and halo-model contributions, reduced shear and source magnification, among others. In addition, we also test the robustness of the results directly on the data by using two independent lens galaxy samples, the so-called \textsc{redMaGiC} sample \citep{y3-galaxyclustering}
and a magnitude-limited sample, \textsc{MagLim} \citep{y3-2x2maglimforecast}, which demonstrates that the lensing ratios information is robust against non-linear small-scale information characterizing the galaxy-matter connection. We also use lensing ratios constructed from large-scale information to further validate the small-scales ratios in the data. After thoroughly validating the shear-ratio likelihood by itself (SR), we proceed to combine it with other 2pt functions and study the improvements it provides in the constraints, using first simulated data and then DES Y3 data. We find SR to provide significant improvements in cosmological constraints, especially for the combination with cosmic shear, due to the information SR provides on IA. The DES Y3 cosmic shear results are described in two companion papers \citep*{y3-cosmicshear1,y3-cosmicshear2}, the results from galaxy clustering and galaxy-galaxy lensing in \citet*{y3-2x2ptmagnification,  y3-2x2ptbiasmodelling,y3-2x2ptaltlensresults} and the combination of all probes in \citet*{y3-3x2ptkp}.

The paper is organized as follows. Section \ref{sec:data} describes the data sets used in this work. In Section \ref{sec:modeling} we detail the modeling of the ratios and the scheme used to do parameter inference using that model. The ratio measurement procedure is described in Section \ref{sec:measurement}. The validation of the model is presented in Section \ref{sec:model_validation}. In Section \ref{sec:combination} we explore the constraining power of the lensing ratios when combined with other probes using simulated data. Finally, in Section \ref{sec:results}, we  apply the methodology to DES Y3 data and present the final results. We summarize and conclude in Section \ref{sec:conclusions}.

\section{Data and simulations}
\label{sec:data}

\begin{figure}
 \centering
 \includegraphics[width=0.45\textwidth]{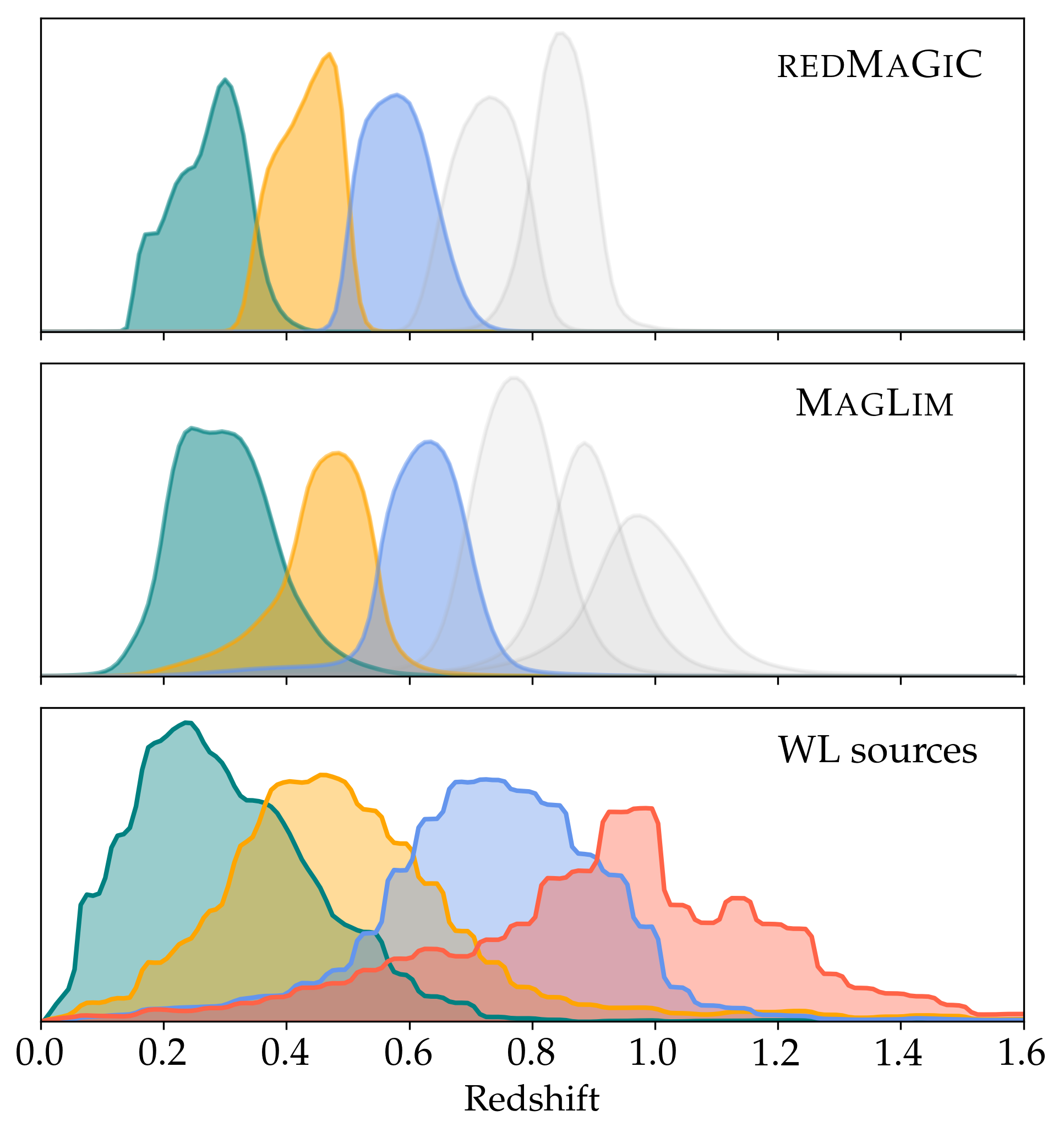}
 \caption{(Top panel): Redshift distributions of \textsc{redMaGiC} lens galaxies divided in five redshift bins. The first three redshift bins are used for the shear ratio analysis in this work, while the two highest-redshift ones (in gray) are not used. The $n(z)$s are obtained by stacking individual $p(z)$ distributions for each galaxy, as computed by the \textsc{redMaGiC} algorithm, and validated using clustering cross-correlations in \citet{y3-lenswz}. (Middle panel): Same as above but for the \textsc{MagLim} lens galaxy sample. The redshift distributions come from the DNF (Directional Neighbourhood Fitting) photometric redshift algorithm \citep{de2016dnf,y3-2x2ptaltlensresults}.  (Bottom panel): The same, but for the weak lensing source galaxies, using the \textsc{Metacalibration} sample. In this case the redshift distributions come from the SOMPZ and WZ methods, described in \citet*{y3-sompz} and \citet*{y3-sourcewz}. }
 \label{fig:nzs}
 \end{figure}

DES is a photometric survey that covers about one 
quarter of the southern sky (5000 sq.~deg.), imaging galaxies in 5 broadband filters ($grizY$) using the Dark Energy Camera \citep{Flaugher2015,DES2016}. In this work we use data from the first three years of observations (from August 2013 to February 2016, hereafter just Y3), which reaches a limiting magnitude ($S/N = 10$) of $\approx 23$ in the $i$-band (with a mean of 4 exposures out of the planned 10 for the full survey), and covers an area of approximately 4100 sq.~deg. The data is processed using the DESDM pipeline presented in \citet{Morganson2018}. For a detailed description of the DES Y3 Gold data sample, see \citet*{y3-gold}. Next we describe the lens and source galaxy samples used in this work. Their corresponding redshift distribution are shown in Figure \ref{fig:nzs}. 

\subsection{Lens samples}

In Table~\ref{tab:samples} we include a summary description for each of the lens samples used in this work, with the number of galaxies in each redshift bin, number density, linear galaxy bias values and lens magnification parameters. 

\subsubsection{The \textsc{redMaGiC} sample}

One of the lens galaxy samples used in this work is a subset of the DES Y3 Gold Catalog selected by \textsc{redMaGiC} \citep{Rozo2015}, which is an algorithm designed to define a sample of luminous red galaxies (LRGs) with high quality photometric redshift estimates. It selects galaxies above some luminosity threshold based on how well they fit a red sequence template, calibrated using the redMaPPer cluster finder \citep{Rykoff2014,Rykoff2016} and a subset of galaxies with spectroscopically verified redshifts. The cutoff in the goodness of fit to the red sequence is imposed as a function of redshift and adjusted such that a constant comoving number density of galaxies is maintained.

In DES Y3 \textsc{redMaGiC} galaxies are used as a lens sample in the clustering and galaxy-galaxy lensing parts of the 3$\times$2pt cosmological analysis \citep{y3-galaxyclustering, y3-gglensing}. In this work we utilize a subset of the samples used in those analyses, in particular the galaxies with redshifts $z<0.65$, split into three redshift bins (see Figure \ref{fig:nzs}). The redshift calibration of this sample is performed using clustering cross-correlations, and is described in detail in \citet{y3-lenswz}. A catalog of random points for \textsc{redMaGiC} galaxies is generated uniformly over the footprint, and then weights are assigned to \textsc{redMaGiC} galaxies such that spurious correlations with observational systematics are cancelled. The methodology used to assign weights is described in \citet{y3-galaxyclustering}.

\subsubsection{The Magnitude-limited sample}

We use a second lens galaxy selection, which differs from \textsc{redMaGiC} in terms of number density and photometric redshift accuracy: the \textsc{MagLim} sample. In this sample, galaxies are selected with a magnitude cut that evolves linearly with the photometric redshift estimate: $i < a z_{\rm phot} + b$. The optimization of this selection, using the DNF photometric redshift estimates, yields $a=4.0$ and $b=18$. This optimization was performed taking into account the trade-off between number density and photometric redshift accuracy, propagating this to its impact in terms of cosmological constraints obtained from galaxy clustering and galaxy-galaxy lensing in \citet{y3-2x2maglimforecast}. Effectively, this selects brighter galaxies at low redshift while including fainter galaxies as redshift increases.  Additionally,  we apply a lower cut to remove the most luminous objects, imposing $i > 17.5$. The \textsc{MagLim} sample has a galaxy number density of more than four times that of the \textsc{redMaGiC} sample but the redshift distributions are $\sim30\%$ wider on average. This sample is split into 6 redshift bins, but in this paper we only use the first three of them. The characteristics of these three redshift bins are defined in Table~\ref{tab:samples}. The redshift binning was chosen to minimize the overlap in the redshift distributions, which is also calibrated using clustering redshifts in \citet{y3-lenswz}. \citet{y3-2x2maglimforecast} showed that changing the redshift binning does not impact the cosmological constraints. See also \citet{y3-2x2ptaltlensresults} for more details on this sample.

\begin{table}
\centering
\pmb{\textsc{redMaGiC} lens sample} \\
\vspace{2mm}
\setlength{\tabcolsep}{5pt}
\begin{tabular}{c|c|c|c|c}
\textbf{Redshift bin}  &   \textbf{$N^i_\text{gal}$} &  \textbf{$n^i_\text{gal}$ } & \textbf{$b^i$} & \textbf{$\alpha^i$} \\
\rule{0pt}{3ex} 
$0.15 < z < 0.35$ & 330243 &  0.022141 & 1.74  $\pm$ 0.12 & 1.31  \\
$0.35 < z < 0.50$ & 571551 &  0.038319 & 1.82  $\pm$ 0.11 & -0.52 \\
$0.50 < z < 0.65$ & 872611 &  0.058504 & 1.92  $\pm$ 0.11 & 0.34 \\
\end{tabular}
\\
\vspace{5mm}
\pmb{\textsc{MagLim} lens sample} \\
\vspace{2mm}
\begin{tabular}{c|c|c|c|c}
\textbf{Redshift bin} &   \textbf{$N^i_\text{gal}$} &  \textbf{$n^i_\text{gal}$ } & \textbf{$b^i$} & \textbf{$\alpha^i$} \\
\rule{0pt}{3ex} 
$0.20 < z < 0.40$ & 2236473 &  0.1499 & 1.49  $\pm$ 0.10 & 1.21 \\
$0.40 < z < 0.55$ & 1599500 &  0.1072 & 1.69  $\pm$ 0.11 & 1.15 \\
$0.55 < z < 0.70$ & 1627413 &  0.1091 & 1.90  $\pm$ 0.12 & 1.88 \\
\end{tabular}
\\
\vspace{5mm}
\pmb{\textsc{Metacalibration} source sample} \\
\vspace{2mm}
\begin{tabular}{c|c|c|c|c}
\textbf{Redshift bin} &   \textbf{$N^j_\text{gal}$} &  \textbf{$n^j_\text{gal}$ } & \textbf{$\sigma_\epsilon^j$} & \textbf{$\alpha^j$}\\
\rule{0pt}{3ex} 
1 & 24940465 &  1.476 & 0.243 & 0.335 \\
2 & 25280405 &  1.479 & 0.262 & 0.685 \\
3 & 24891859 &  1.484 & 0.259 & 0.993 \\
4 & 25091297 &  1.461 & 0.301 & 1.458 \\
\end{tabular}

\caption{Summary description for each of the samples used in this work. $N_\text{gal}$ is the number of galaxies in each redshift bin, $n_\text{gal}$ is the effective number density in units of gal/arcmin$^{2}$ (including the weights for each sample), $b^i$ is the mean linear galaxy bias from the 3$\times$2pt combination, the \textbf{$\alpha$}'s are the magnification parameters as measured in \citet{y3-2x2ptmagnification} and $\sigma_\epsilon^j$ is the weighted standard deviation of the ellipticity for a single component as computed in \citet*{y3-shapecatalog}.}
\label{tab:samples}
\end{table}

\subsection{Source sample}\label{sec:source_sample}

The DES Y3 source galaxy sample, described in \citet*{y3-shapecatalog}, comprises a subset of the DES Y3 Gold sample.  It is based on \textsc{Metacalibration} \citep{Huff2017,Sheldon2017}, which is a method developed to accurately measure weak lensing shear using only the available imaging data, without need for prior information about galaxy properties or calibration from simulations. The method involves distorting the image with a small known shear, and calculating the response of a shear estimator to that applied shear. This technique
can be applied to any shear estimation code provided it fulfills certain requirements. For this work, it has been applied to the \textsc{ngmix} shear pipeline \citet{Sheldon2014}, which fits a Gaussian model simultaneously in the $riz$ bands to measure the ellipticities of the galaxies. The details of this implementation can be found in \citet*{y3-shapecatalog}.  

The redshift calibration of the source sample has been performed using the Self Organizing Maps Photometric Redshifts (SOMPZ, \citealt*{y3-sompz}) and the clustering cross-correlation (WZ, \citealt*{y3-sourcewz}) method.  The SOMPZ scheme uses information from the DES Deep Fields \citep*{y3-deepfields} and connects it to the wide survey by using the Balrog transfer function \citep*{y3-balrog}. Using the that method, the source sample is split into four redshift bins (Figure \ref{fig:nzs}), and the scheme provides a set of source redshift distributions, including the uncertainty from sample variance, flux measurements, etc. The WZ method uses the cross-correlations of the positions of the source sample with the positions of the \textsc{redMaGiC} galaxies, narrowly binned in redshift. For its application, samples are drawn from the posterior distribution of redshift distributions for all bins conditioned on both the SOMPZ photometric data and the WZ clustering data. In addition, validation of the shape catalog uncertainties, and the connection to uncertainties in the associated redshift distributions has been developed in detail in \citet{y3-imagesims}, using realistic image simulations. For this work we will employ the shear catalog and use the results from these analyses as priors on source multiplicative biases and redshift calibration.

In Table~\ref{tab:samples} we include the number of galaxies in each redshift bin as well as the number density, shape noise and source magnification parameters. 

\subsection{$N$-body simulations}

In this work we use $N$-body simulations to recreate an end-to-end analysis and validate our methodology. For this we use the Buzzard simulations described in Sec.~\ref{sec:buzzard}. We also use the MICE2 simulations to validate our small scale \texttt{Halofit} modeling with a Halo Occupation Distribution (HOD) model. We describe the MICE2 simulation in Sec.~\ref{sec:mice}. 

\subsubsection{The Buzzard v2.0 $N$-body simulations}
\label{sec:buzzard}

Buzzard v2.0 \citep{y3-simvalidation} is a suite of 18 simulated galaxy catalogs built on $N$-body lightcone simulations that have been endowed with a number of DES Y3 specific survey characteristics. Each pair of 2 Y3  simulations are produced from a set of 3 independent $N$-body lightcones with mass resolutions of $3.3\times10^{10},\, 1.6\times10^{11},\, 5.9\times10^{11}\, h^{-1}M_{\odot}$, and simulated volumes of $1.05,\, 2.6 \textrm{ and } 4.0\, (h^{-3}\, \rm Gpc^3)$. Galaxies are included in these simulations using the \textsc{Addgals} model \citep{Wechsler2021, DeRose2021}. \textsc{Addgals} makes use of the relationship, $P(\delta_{R}|M_r)$, between a local density proxy, $\delta_{R}$, and absolute magnitude $M_r$ measured from a high resolution sub--halo abundance matching (SHAM) model in order to populate galaxies into these lightcone simulations. This model reproduces the absolute--magnitude--dependent clustering of the SHAM.

The \textsc{Calclens} algorithm is used to ray-trace the simulations, using a spherical-harmonic transform (SHT) based Poisson solver \citep{Becker2013}. A $N_{\rm side}=8192$ \textsc{HealPix} grid is used to perform the SHTs. \textsc{Calclens} computes the lensing distortion tensor at each galaxy position, and uses this quantity to deflect galaxy angular positions, shear galaxy intrinsic ellipticities, including effects of reduced shear, and magnify photometry and shapes. Convergence tests have shown that resolution effects are negligible in relevant lensing quantities on the scales used for this analysis \citep{DeRose2019}.

We apply a photometric error model based on DES Y3 data estimates in order to add realistic wide field photometric noise to our simulations. A lens galaxy sample is selected from our simulations by applying the \textsc{redMaGiC} galaxy selection with the configuration described in \citet{y3-galaxyclustering}. A weak-lensing source galaxy selection is performed by selecting on PSF-convolved sizes and $i$-band signal-to-noise in a manner that matches the measured non-tomographic source number density in the DES Y3 \textsc{metacalibration} source catalog. SOMPZ redshift estimation is used in the simulations in order to place galaxies into four source redshift bins. The shape noise per redshift bin is also matched to that measured from the \textsc{metacalibration} catalog. Two-point functions are measured in the Buzzard v2.0 simulations with the same code used for the Y3 data. \textsc{metacalibration} responses and inverse variance weights are set equal to 1 for all galaxies, because our simulations do not include these values. Weights for the simulated lens galaxy sample are assigned using the same algorithm used in the DES Y3 data.

\subsubsection{The MICE2 $N$-body simulation}\label{sec:mice}

We use DES-like mock galaxy catalogs from the MICE
simulation suite in this analysis. The MICE Grand Challenge simulation (MICE-GC) is an $N$-body simulation
run in a cube with side-length 3 Gpc/$h$ with $4096^3$ particles using the Gadget-2 code \citep{Springel_2005} with mass resolution of $2.93\times 1010 M_\odot/h$. Halos are identified using a Friends-of-Friends algorithm with linking length 0.2. For further details about this simulation, see \citet{Fosalba2015a}.
These halos are then populated with galaxies using a
hybrid sub-halo abundance matching plus halo occupation distribution (HOD) approach, as detailed in \citet{Carretero2014}. These methods are designed to match
the joint distributions of luminosity, $g - r$ color, and
clustering amplitude observed in SDSS \citep{Zehavi_2005}. The construction of the halo and galaxy catalogs is described in \citet{Crocce2015a}. MICE assumes a flat $\Lambda$CDM cosmological model with $h = 0.7$, $\Omega_m = 0.25$, $\Omega_b = 0.044$ and $\sigma_8 = 0.8$, and it populates one octant of the sky (5156 sq. degrees), which is comparable to the sky area of DES Y3 data.

To validate our small scale \texttt{Halofit} modeling in Sec~\ref{sec:hod}, testing it against an HOD model with parameters measured from MICE2, we use a DES-like lightcone catalog of \textsc{redMaGiC} galaxies matching the properties of DES Y3 data, including lens magnification.

\section{Modeling of the ratios}
\label{sec:modeling}

In this section we describe how we model the ratios of tangential shear measurements and why it is possible to model them to significantly smaller scales than the tangential shear quantity.

\subsection{The idea: Geometrical ratios}
\label{sec:geometrical_model}

When we take ratios of tangential shear measurements around the same lens sample, the dependence on the matter power spectrum and galaxy bias cancels for the most part, canceling exactly if the lens sample is infinitely narrow in redshift. In this approximation the ratios can be modelled independently of scale, and they depend only on the geometry of the Universe. As we will see now, this fact allows us to model ratios of tangential shear measurements down to significantly smaller scales than what is typically used for the tangential shear measurements themselves. For instance, in the case of the DES Y3 3$\times$2pt cosmological analysis, scales below 6 Mpc/$h$ are discarded for the galaxy-galaxy lensing probe due to our inability to accurately model the (non-linear) matter power spectrum, the galaxy bias, baryonic effects, etc. In order to see why these dependencies may cancel out in the ratios, it is useful to first express the tangential shear $\gamma_t$ in terms of the excess surface mass density $\Delta \Sigma$:
\begin{equation}\label{eq:gammat_delta_sigma}
\gamma_t = \frac{\Delta \Sigma}{\Sigma_\mathrm{crit}},
\end{equation}
where the lensing strength $\Sigma_{\mathrm{crit}}^{-1}$ is a geometrical factor that, for a single lens-source pair, depends on the angular diameter distance to the lens $D_{\rm l}$, the source $D_{\rm s}$ and the relative distance between them $D_{\rm ls}$:
\begin{equation}\label{eq:inverse_sigma_crit}
\Sigma_{\mathrm{crit}}^{-1} (z_{\rm l}, z_{\rm s}) = \frac{4\pi G}{c^2} \frac{D_{\rm ls} \, D_{\rm l}}{D_{\rm s}},
\end{equation}
with $\Sigma_{\mathrm{crit}}^{-1}(z_l,z_s)=0$ for $z_s<z_l$, and where $z_l$ and $z_s$ are the lens and source galaxy redshifts, respectively. For a single lens-source pair, Eq.~(\ref{eq:gammat_delta_sigma}) is exact and can be used to see that if one takes the ratio of two tangential shear measurements sharing the same lens with two different sources, $\Delta \Sigma$ cancels since it is a property of the lens only (see \citet{Bartelmann2001} for a review), and we are left with a ratio of geometrical factors:
\begin{equation}\label{eq:ratios_narrow_lens}
\frac{\gamma_{t}^{l,s_i}}{\gamma_{t}^{l,s_j}} = \frac{\Sigma_{\mathrm{crit}}^{-1} (z_l, z_{s_i})}{\Sigma_{\mathrm{crit}}^{-1} (z_l, z_{s_j})}.
\end{equation} 
This means that ratios defined in this way will depend on the redshift of the lens and source galaxies, as well as on the cosmological parameters needed to compute each of the angular diameter distances involved in  Eq.~(\ref{eq:inverse_sigma_crit}), through the distance–redshift relation.

So far we have only been considering a single lens-source pair. For a tangential shear measurement involving a sample of lens galaxies with redshift distribution $n_l(z)$ and a sample of source galaxies with $n_s(z)$, which may also overlap, we can generalize Eq.~(\ref{eq:ratios_narrow_lens}) by defining an effective $\Sigma^{-1}_{\mathrm{crit}}$ integrating over the corresponding redshift distributions. For a given lens bin $i$ and source bin $j$, it can be expressed as:
\begin{equation}\label{eq:eff_inverse_sigma_crit}
\Sigma_{\mathrm{crit},\mathrm{eff}}^{-1\ i,j} = \int_0^{z_l^\text{max}} dz_l  \int_0^{z_s^\text{max}} dz_s \, n_l^i(z_l) \, n_s^j(z_s) \, \Sigma_{\mathrm{crit}}^{-1}(z_l, z_s).
\end{equation}
Then, the generalized version of Eq.~(\ref{eq:ratios_narrow_lens}) becomes:
\begin{equation}\label{eq:ratios_eff_sigma_crit}
\frac{\gamma_{t}^{l,s_i}}{\gamma_{t}^{l,s_j}} \simeq \frac{\Sigma_{\mathrm{crit},\mathrm{eff}}^{-1 \ l,s_i}}{\Sigma_{\mathrm{crit},\mathrm{eff}}^{-1 \ l, s_j}}.
\end{equation} 

In this equation it  becomes apparent that the main dependency of the ratios is on the redshift distributions of both the lens and the source samples.  Eq.~(\ref{eq:ratios_narrow_lens}) is only exact if the lens sample is infinitely narrow in redshift and a good approximation if the lens sample is narrow \textit{enough}. This approximation is what was used in the DES Y1 shear-ratio analysis \citep*{Y1GGL} to model the ratios. In this work we will go one step further and we will not use the narrow-lens bin approximation. Instead, we will use a full modeling of the ratios adopting the tangential shear model used in the DES Y3 3$\times$2pt analysis, which includes explicit modeling of other effects such as lens magnification, intrinsic alignments and multiplicative shear biases, which will also play a role in the ratios. Next we describe in detail the full modeling of the ratios we use in this work.

\subsection{The full model}
\label{sec:model}

\begin{figure}
 \centering
 \includegraphics[width=0.48\textwidth]{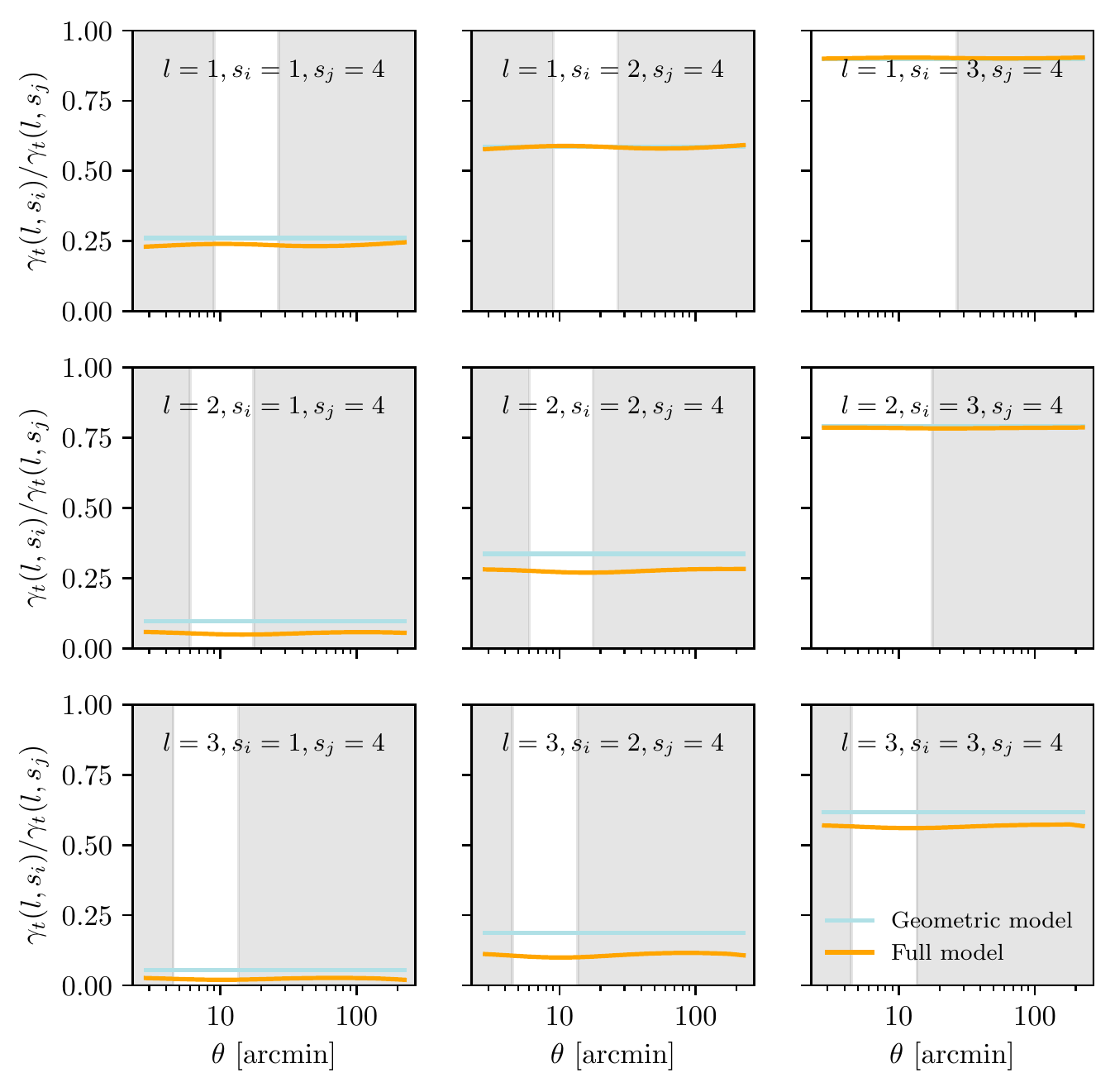}
 \caption{Lensing ratios using the full model of the ratios we use in this work as a function of scale evaluated at the best-fit values of the 3$\times$2pt analysis (see Sec.~\ref{sec:model}) compared with the purely geometrical model used in previous shear-ratio analyses until this date, which is scale independent (see Sec.~\ref{sec:geometrical_model}). We can appreciate the geometrical component still dominates the modeling of the ratios but small but significant deviations are found when comparing with the full modeling. 
The unshaded regions correspond to the ``small scales'' we use in this analysis, which are adding extra information below the scales used in the 3$\times$2pt cosmological analysis for the galaxy-galaxy lensing probe. The grey shaded regions are not used for the fiducial ratios in this work.}
 \label{fig:ratios_scale}
 \end{figure}

Ratios of tangential shear measurements are the main probe used in this work. In this section we will describe how we model it for our fiducial case, including the integrals of the power spectrum over the lens bins range, where we do not use the narrow lens bin approximation, and with contributions from lens magnification and intrinsic alignments. 
The model of the ratio for the lens redshift bin $i$ between source redshift bins $j$ and $k$ can be expressed as:
\begin{equation}
\label{eq:ratio}
r^{(l_i,s_j,s_k)} \equiv \left \langle \frac{   \gamma_t^{l_i,s_j} (\theta)}{ \gamma_t^{l_i,s_k} (\theta) } \right \rangle_\theta =  \left \langle r^{(l_i,s_j,s_k)} (\theta) \right \rangle_\theta ,
\end{equation}
where the averaging over the different angular bins is performed as detailed in Sec.~\ref{sec:measurement}, in the same way as we do it for the measurement. To model each tangential shear quantity in the ratio, we use exactly the same model used for the galaxy-galaxy lensing probe in the DES Y3 3$\times$2pt cosmological analysis, which we will summarize in this section and for which further details can be found in \citet{y3-generalmethods} and \citet{y3-gglensing}. Also, in Fig.~\ref{fig:ratios_scale} we show the full modeling of the ratios as a function of scale, before performing the angular averaging. We compare it with the purely geometrical modeling described in the previous section, and find that, even though the geometrical part component continues to be the dominant component needed to model the ratios, other contributions become significant for some of the lens-source bins combinations and thus the full modeling is needed.  

The tangential shear two-point correlation function for each angular bin can be expressed as a transformation of the galaxy-matter angular cross-power spectrum $C_{gm}(\ell)$, which in this work we perform using the curved sky projection:
\begin{equation}\label{eq:curved_sky}
\gamma_t^{ij}(\theta) = (1+m^j)\sum_\ell \frac{2\ell+1}{4\pi\ell(\ell+1)} \overline{P^2_\ell}\left(\theta_{\rm min},\theta_{\rm max}\right) \, C^{ij}_{gm, \text{tot}}(\ell),
\end{equation}
for a lens redshift bin $i$ and a source redshift bin $j$, where $\overline{P^2_\ell}\left(\theta_{\rm min},\theta_{\rm max}\right) $ is the bin-averaged associated Legendre polynomial within an angular bin $[\theta_{\rm min},\theta_{\rm max}]$, defined in \citet{y3-gglensing}. $m^j$ are free parameters that account for a multiplicative uncertainty on the shape measurements. The total angular cross-power spectrum $C^{ij}_{gm, \text{tot}}$ in the equation above includes terms from  Intrinsic Alignments (IA), lens magnification, and cross terms between the two effects:
\begin{equation}
C^{ij}_{gm, \text{tot}} = C^{ij}_{gm} + C^{ij}_{gm, \text{IA}} + C^{ij}_{gm, \text{lens mag}} + C^{ij}_{gm, \text{IA x lens mag}}.
\end{equation}
The main angular cross-power spectrum can be written as this projection of the 3D galaxy-matter power spectrum $P_{gm}$, using Limber's approximation \citep{Limber53, Limber_LoVerde2008} and assuming a flat Universe cosmology:
\begin{equation}\label{eq:C_gm}
 C_{gm}^{ij}(\ell) = \frac{3H^2_0 \Omega_m }{2c^2} 
    \int d\chi N_l^i(\chi) \frac{g^j(\chi)}{a(\chi)\,  \chi}
    P_{gm}\scriptstyle{\left(\frac{\ell+1/2}{\chi},z(\chi)\right)},
\end{equation}
where
\begin{equation}\label{eq:N_l}
N_l^i(\chi) = \frac{n^i_l\,(z-\Delta z^i_l)}{\bar{n}^i_l}\frac{dz}{d\chi},
\end{equation}
with $\Delta z^i_l$ accounting for the uncertainty on the mean redshift of the lens redshift distributions. For the \textsc{MagLim} sample we also marginalize over the width of the lens redshift distributions, introducing the parameters $\sigma_{z_l}^i$, one for each lens redshift bin (see \citealt{y3-2x2ptaltlensresults,y3-lenswz} for additional details about the introduction of the width parameterization). In the equations above, $k$ is the 3D wavenumber, $\ell$ is the 2D multipole moment, $\chi$ is the comoving distance to redshift $z$, $a$ is the scale factor, $n^i_l$ is the lens redshift distribution, $\bar{n}^i_l$ is the mean number density of the lens galaxies and $g(\chi)$ is the lensing efficiency kernel: 
\begin{equation}\label{eq:lensing_efficiency}
 g(\chi) = \int_\chi^{\chi_\text{lim}} d \chi' N^j_s(\chi') \frac{\chi'- \chi}{\chi'}
\end{equation}
with $N^j_s(\chi')$ being analogously defined for the source galaxies as in Eq.~(\ref{eq:N_l}) for the lens galaxies, introducing the source redshift uncertainty parameters $\Delta z^j_s$. $\chi_\mathrm{lim}$ is the limiting comoving distance of the source galaxy sample. Also, we want to relate the galaxy-matter power spectrum to the matter power spectrum for all the terms above. In our fiducial model we assume that lens galaxies trace the mass distribution following a simple linear biasing model ($\delta_g = b \:\delta_m$). The galaxy-matter power spectrum relates to the matter power spectrum by a multiplicative galaxy bias factor:
\begin{equation} \label{eq:linearbias}
 P^{ij}_{gm} = b^{i}  P^{ij}_{mm},
\end{equation}
even though the galaxy bias mostly cancels in the lensing ratios. We find the lensing ratios to have significant dependence on the IA and lens magnification terms but almost no sensitivity to the galaxy bias model. We compute the non-linear matter power spectrum $P_{mm}$ using the \citet{Takahashi_2012} version of \texttt{Halofit} and the linear power spectrum with \texttt{CAMB}\footnote{https://camb.info/}. 
To compute the theoretical modelling in this study, we use the \textsc{CosmoSIS} framework \citep{Zuntz_2015}. 

Below we briefly describe the other terms included in our fiducial model.

\paragraph*{Lens magnification} Lens magnification is the effect of magnification applied to the lens galaxy sample by the structure that is between the lens galaxies and the observer. The lens magnification angular cross-power spectrum can be written as:
\begin{equation}
C^{ij}_{gm, \text{lens mag}}  = 2 (\alpha^i -1) \, C_{mm}^{ij}(\ell)
\end{equation}
where $\alpha^i$ is a parameter that depends on the properties of the lens sample and has been measured in \citet{y3-2x2ptmagnification} for the DES Y3 lens samples within the 3$\times$2pt analysis. The measured values can be seen in Table~\ref{tab:samples}. $C_{mm}^{ij}(\ell)$ is the convergence power spectrum between the lens and source distributions, as defined in \citet{y3-2x2ptmagnification}. 

\paragraph*{Intrinsic Alignments} The orientation of the source galaxies is correlated with the underlying large-scale structure, and therefore with the lenses tracing this structure. This effect is only present in galaxy-galaxy lensing measurements if the lens and source galaxies overlap in redshift. To take it into account, we employ the TATT (Tidal Alignment and Tidal Torquing)  model \citep{Blazek_2019} which is an extension of the NLA (Non-linear alignment)  model \citep{Hirata2004}. Then, the IA term is:
\begin{equation}
    C_{IA}^{ij}(\ell) = \int d\chi \frac{N_l^i(\chi)\, N_{s}^j(\chi)}{\chi^2}P_{gI}\left(k = \frac{\ell+1/2}{\chi},z(\chi)\right)\,,
\end{equation}
where $P_{gI} = b P_{GI}$, with $b$ being the linear bias of the lens galaxies. $P_{GI}$ is model dependent and in the TATT model is given by:
\begin{align}
P_{GI} = a_1(z) P_{mm} + a_{1\delta}(z) P_{0|0E}
+ a_2 (z) P_{0|E2},
\end{align}
where the full expressions for the power spectra  of the second and third term in the can be found in \citet{Blazek_2019} (see equations 37-39 and their appendix A). The other parameters are defined as:
\begin{equation}\label{eq:tatt_c1}
    a_1(z) = -A_1 \bar{C}_{1} \frac{\rho_{\rm crit}\Omega_{\rm m}}{D(z)} \left(\frac{1+z}{1+z_{0}}\right)^{\eta_1}
\end{equation}
\begin{equation}\label{eq:tatt_c2}
    a_2(z) = 5 A_2 \bar{C}_{1} \frac{\rho_{\rm crit}\Omega_{\rm m}}{D^2(z)} \left(\frac{1+z}{1+z_{0}}\right)^{\eta_2}
\end{equation}
\begin{equation}\label{eq:b_ta}
a_{1\delta} (z) = b_{\mathrm{TA}} a_1 (z),
\end{equation}
where $\bar{C}_1$ is a normalisation constant, by convention fixed at a value $\bar{C}_1=5\times10^{-14}M_\odot h^{-2} \mathrm{Mpc}^2$, obtained from SuperCOSMOS (see \citealt{brown02}). The denominator $z_{0}$ is a pivot redshift, which we fix to the value 0.62. Finally, the dimensionless amplitudes $(a_1, a_2)$, the power law indices $(\eta_1,\eta_2)$ and the $b_{\text{TA}}$ parameter (which accounts for the fact that the shape field is preferentially sampled in overdense regions) are the 5 free parameters of our TATT model.

\paragraph*{Lens magnification cross Intrinsic Alignments term} There is also the contribution from the correlation between lens magnification and source intrinsic alignments, which is included in our fiducial model:
\begin{equation}
    C_{mI}^{ij}(\ell) = \int d\chi \frac{q_l^i(\chi)\, N_{s}^j(\chi)}{\chi^2}P_{mI}\left(k = \frac{\ell+1/2}{\chi},z(\chi)\right)\,,
\end{equation}
where $P_{mI} = P_{GI}$.

\subsubsection{Parameters of the model}

Next we will describe the different dependencies of the modeling of the ratios. In most cases, such dependencies will be described by parameters in our model (listed below), some of which will have Gaussian priors associated with them.

\begin{itemize}
    \item \textbf{Cosmological parameters (6 or 7)}: 6 for $\Lambda$CDM, which are $\Omega_m$, $H_0$, $\Omega_b$, $n_s$, $A_s$ (or $\sigma_8$\footnote{We sample our parameter space with $A_s$, and convert to $\sigma_8$ at each step of the chain to get the posterior of $\sigma_8$. We also use the parameter $S_8$, which is a quantity well constrained by weak lensing data, defined here as $S_8 = \sigma_8(\Omega_m/0.3)^{0.5}$.}) and $\Omega_{\nu} h^2$. For $w$CDM, there is an additional parameter, $w$, that governs the equation of state of dark energy. Also, in our model we are assuming 3 species of massive neutrinos, following \citet{y3-3x2ptkp}, and a flat geometry of the Universe ($\Omega_k=0)$. 
    \item \textbf{Source redshifts parameters}: In order to characterize the uncertainties, we allow for an independent shift $\Delta z^j$ in each of the measured source redshift distributions. Priors for these parameters have been obtained in \citet*{y3-sompz, y3-sourcewz}. Additional validation with respect to marginalizing over the shape of the source redshift distributions is provided in \citet{y3-hyperrank} using the \texttt{Hyperrank} method.
    \item \textbf{Lens redshift parameters}: We allow for independent shifts in the mean redshift of the distributions, $\Delta z^i$, one per each $i$ lens redshift bin, as defined in Eq.~(\ref{eq:N_l}). For the \textsc{MagLim} sample there are additional parameters to marginalize over: the width of the redshift distributions $\sigma_{z^i}$. That is because the width of the distributions is more uncertain in the \textsc{MagLim} case. Priors for these parameters have been obtained in \citet{y3-lenswz}.
    \item \textbf{Multiplicative shear bias parameters}: We allow for a multiplicative change on the shear calibration of the source samples using $m^j$, one per each $j$ source redshift bin. Priors for these parameters have been obtained in \citet{y3-imagesims}.
    \item \textbf{Lens magnification parameters}: They describe the sign and amplitude of the lens magnification effect. We denote them by $\alpha^i$, one per each $i$ lens redshift bin. These parameters have been computed in \citet{y3-2x2ptmagnification} and are fixed in our analysis as well as in the 3$\times$2pt analysis.
    \item \textbf{(Linear) Galaxy bias parameters}: They model the relation between the underlying dark matter density field and the galaxy density field: $b^i$, one per each $i$ lens redshift bin, since we assume linear galaxy bias in this analysis. 
    \item \textbf{Intrinsic Alignment (IA) parameters}: Our fiducial IA model is the TATT model, which has 5 parameters: two amplitudes governing the strength of the alignment for the tidal and for the torque par, respectively, $a_1$, $a_2$, two parameters modeling the dependence of each of the amplitudes in redshift, $\alpha_1$, $\alpha_2$, and $b_{\mathrm{TA}}$, describing the galaxy bias of the source sample.   
\end{itemize}

\subsubsection{Different run configurations} \label{sec:model_dv_schemes}
Now we have listed all the dependencies of the model used to describe the ratios throughout this paper. Across the paper, however, we will perform different tests using the ratios, freeing different parameters in each case. We consider three main different scenarios, and for each scenario we use different measurements and allow different parameters to vary. The tests will be described in more detail as they appear in the paper, but here we list these distinct scenarios and the modeling choices adopted in each of them:
\begin{enumerate}
    \item \textbf{Shear-ratio only (SR)}: In this case, the data vector consists of small-scale shear-ratio measurements only (see Sec.~\ref{sec:small-large-ratios} for the definition of scales used). The model has 19 free parameters for the \textsc{redMaGiC} sample: 3 lens redshift parameters, 4 source redshift parameters, 4 multiplicative shear bias parameters, 3 galaxy bias parameters and 5 IA parameters. For the \textsc{MagLim} sample there are 22 free parameters, with the additional 3 lens redshift parameters describing the width of the distributions. In this case we fix the cosmological parameters since the lensing ratios have been found to be insensitive to cosmology (see Sec.~\ref{sec:cosmology_dependence} for a test of this assumption). 
    \item \textbf{Large-scales shear-ratio only (LS-SR)}: In this case, the data vector consists of large-scale shear-ratio measurements only (see Sec.~\ref{sec:small-large-ratios} for the definition of scales used). The model (and number of free parameters) is the same one as for the small scales lensing ratios scenario. This setup is only used as validation for the small-scale shear-ratio analysis.
    \item \textbf{Shear-ratio + 3$\times$2pt (SR + 3$\times$2)}: In this case, the data vector consists of small-scale shear-ratio measurements and the usual 3$\times$2pt data vector, that is, galaxy clustering $w(\theta)$, galaxy-galaxy lensing, $\gamma_t(\theta)$, and cosmic shear, $\xi_+(\theta)$, $\xi_-(\theta)$ measurements, each one with the corresponding scale cuts applied to the DES Y3 3$\times$2pt cosmological analysis. In this case we used exactly the same model as in the 3$\times$2pt cosmological analysis, freeing all the parameters described above, that is, for the \textsc{redMaGiC sample} 29 parameters in total for $\Lambda$CDM, 30 for $w$CDM, and 31 for the \textsc{MagLim} sample for $\Lambda$CDM, 32 for $w$CDM. The only difference between this scenario and the 3$\times$2pt one is the addition of the small-scales lensing ratios measurements in the data vector. 

\end{enumerate}

\subsection{Parameter inference methodology }
\label{sec:parameter_inference_methodology}

In this work we want to use ratios of small-scale galaxy-galaxy lensing measurements around the same lens bins to constrain redshift uncertainties and other systematics or nuisance parameters of our model, as described above. Next we summarize the methodology we utilize to perform such tasks using Bayesian statistics. 

Let's denote the set of measured ratios as $\{ r \}$, and the set of parameters in our model as $\{ M \}$. We want to know the probability of our model parameters given the ratios data. In particular, we are interested in estimating the \textit{posterior} probability distribution function of each parameter in our model ($\{ M \}$) given the ratios data $\{ r \}$, $p(\{ M \}|\{ r \})$. In order to get that posterior probability, we will use Bayes theorem, which relates that posterior distribution to the \textit{likelihood}, $p(\{ r \}|\{ M \})$, computed from the model and the data, and the \textit{prior}, $p(\{ M \})$, which encapsulates \textit{a priori} information we may have on the parameters of our model, via the following relation:
\begin{equation}
p(\{ M \}|\{ r \}) \propto p(\{ r \}|\{ M \}) \: p(\{ M \}).    
\end{equation}


We will use a given set of priors on the model parameters; some of them will be uniform priors in a certain interval, others will be Gaussian priors in the cases where we have more information about the given parameters. For SR, we will assume a Gaussian likelihood, which means that for a given set of parameters in the model ($\{ M \}$), we will compute the corresponding ratios for those model parameters ($\{r\}_M$), and then estimate a $\chi^2$ value between these and the data ratios ($\{r\}$), using a fixed data covariance ($\mathbf{C}$), and then the logarithm of the SR likelihood becomes:

\begin{equation}
    \mathrm{log} \: \mathcal{L}^{\mathrm{SR}} = \mathrm{log} \: p(\{r\}|\vec{M})  = - \frac{1}{2}\chi^2 \:  -\frac{1}{2} \mathrm{log} \: \mathrm{Det} \: \mathbf{C}; 
\end{equation}
\begin{equation}
    \mathrm{with} \; \chi^2 = (\{ r \}-\{ r \}_M)^T \: \mathbf{C}^{-1} \: (\{ r \}-\{ r \}_M).
\end{equation}

This method will provide constraints on the parameters of our model given the measured ratios on the data and a covariance for them. For the fiducial DES Y3 cosmological analysis, this SR likelihood will be used in combination with the likelihood for other 2pt functions such as cosmic shear, galaxy clustering and galaxy-galaxy lensing. Because SR is independent of the other 2pt measurements (see Sec.~\ref{sec:independence_between_small_large}), the likelihoods can be simply combined:
\begin{equation}\label{eq:combined_likelihood}
\mathrm{log} \: \mathcal{L}^{\mathrm{Total}} = \mathrm{log} \: \mathcal{L}^{\mathrm{SR}} + \mathrm{log} \: \mathcal{L}^{\mathrm{2pt}} . 
\end{equation}
The specific details of the parameters and the associated priors used in each test will be described in detail later in the paper, together with the description of the test itself. For MCMC chains, we use \textsc{PolyChord} \citep{polychord} as the fiducial sampler for this paper. We use the following settings for this sampler: \texttt{feedback = 3}, 
\texttt{fast\_fraction = 0.1}, \texttt{live\_points = 500}, \texttt{num\_repeats=60}, \texttt{tolerance=0.1}, \texttt{boost\_posteriors=10.0} for the chains ran on data, and \texttt{live\_points = 250}, \texttt{num\_repeats=30} for chains on simulated data vectors, consistent with \citet{y3-3x2ptkp} and following the guidelines from \citet*{y3-samplers}.

\section{Measurement and covariance of the ratios}
\label{sec:measurement}

In this Section, we describe the measurement and covariance of the ratios, including the choice of scales we use, and we test the robustness of the estimation. The measurement of the ratios is based on the tangential shear measurements presented and validated in \citet{y3-gglensing}, where several measurement tests are performed on the 2pt measurements, such as testing for B-modes, PSF leakage, observing conditions, scale-dependent responses, among others.

\subsection{Methodology}\label{sec:measurement_methodology}

\subsubsection{Lens-source bin combinations}

In this work we use three lens redshift bins, for both lens galaxy samples, \textsc{redMaGiC} and \textsc{MagLim}, and four source redshift bins, as described in Section \ref{sec:data} and depicted in Figure \ref{fig:nzs}. The DES Y3 3$\times$2pt project uses five and six lens bins (Figure \ref{fig:nzs}) for the two lens samples, respectively. In this work we stick to the three lowest redshift lens bins both because they carry the bulk of the total shear ratio S/N and because the impact of lens magnification is much stronger for the highest redshift lens bins, and we choose not to be dominated by lens magnification even though we include it in the modeling, given the uncertainty in the parameters calibrating it. Regarding source redshift bins, we use the four bins utilized in the DES Y3 3$\times$2pt project. 

From the redshift bins described above, we will construct combinations with a given fixed lens bin and two source bins, denoted by the label $(l_i,s_j,s_k)$, where $s_k$ corresponds to the source bin which will sit in the denominator. Then, for each lens bin we can construct three independent ratios, to make a total set of 9 independent ratios, $\{ r^{(l_i,s_j,s_k)} \}$. Note there is not a unique set of independent ratios one can pick. In this work we choose to include the highest S/N tangential shear measurement in the denominator of all of the ratios since that choice will minimize any potential noise bias. In our case the highest S/N tangential shear measurement corresponds to the highest source bin, i.e.~the fourth one, and hence, for a given lens bin $i$ we will use the following three independent ratios {$\{ r^{(l_i,s_1,s_4)} \}, \{ r^{(l_i,s_2,s_4)} \}, \{ r^{(l_i,s_3,s_4)} \}$}. See also Table~\ref{tab:scale_cuts} for a complete list of the ratio combinations we use in this work.

\subsubsection{Small and large scale ratios: choice of scales}
\label{sec:small-large-ratios}

\begin{figure}
 \centering
 \includegraphics[width=0.52\textwidth]{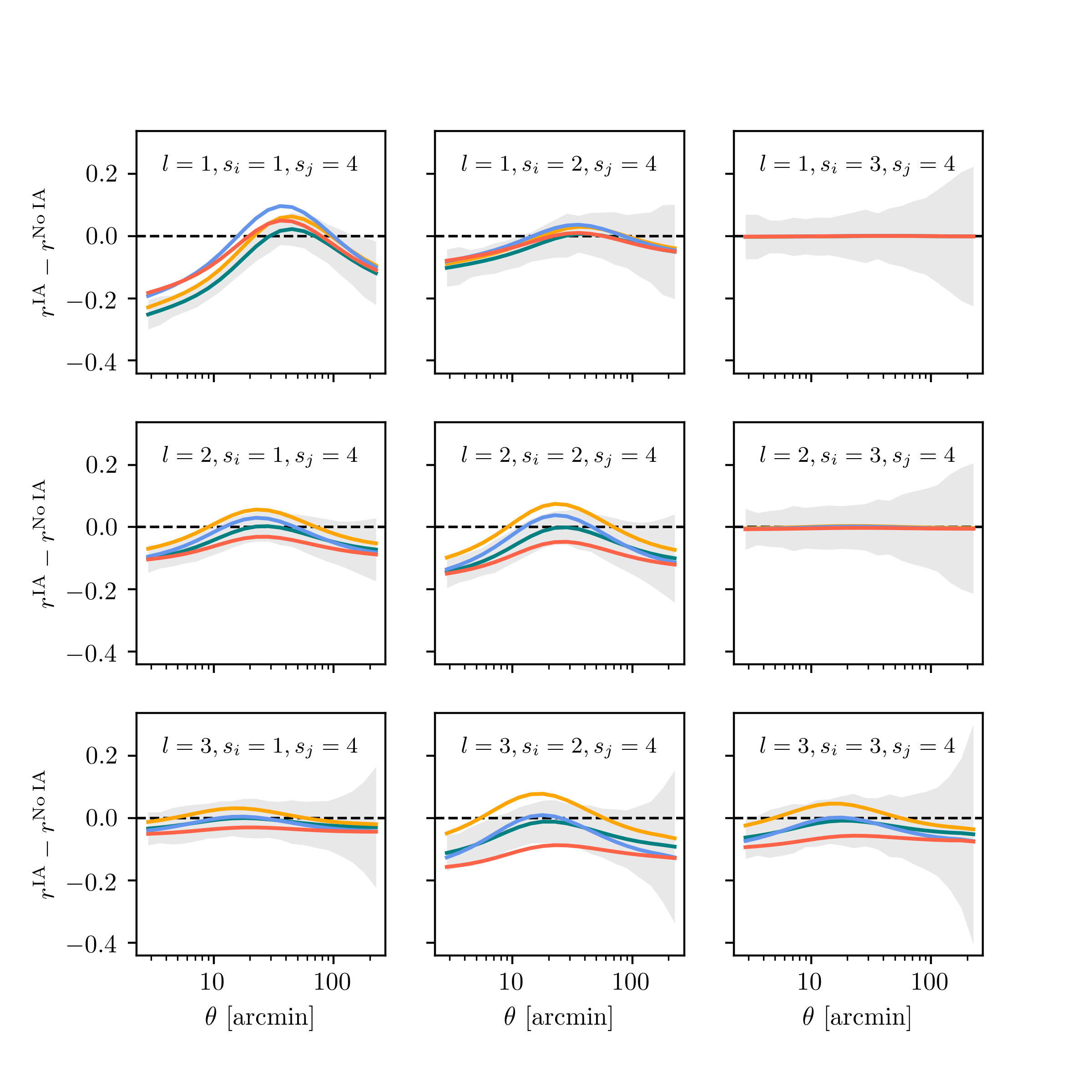}
 \caption{Impact of different IA models (different parameter choices for TATT) on all the lensing ratios considered in this work. We find that for ratios whose modeling is close to a pure geometrical model (Figure \ref{fig:ratios_scale}) the impact of IA is negligible.  The different lines in the plot have different IA parameters in the ranges: $a_1 = [0.5,1], \  a_2 = [-2, -0.8], \ \alpha_1 = [-2.5,0], \alpha_2 = [-4., -1.2], \ b_{TA} = [0.6, 1.2]$. The grey bands show the size of the data uncertainties on the ratios, for reference.}
 \label{fig:ratios_scale_IA}
 \end{figure}

When measuring the lensing ratios, we will be interested in two sets of angular scales, which we label as ``small-scale ratios'' and ``large-scale ratios''. Large-scale ratios are defined to use approximately the same angular scales as the galaxy-galaxy lensing probe in the 3$\times$2pt DES Y3 cosmological analysis, and for that we use scales above 8 $h^{-1}$Mpc and until angular separations of 250 arcmins. In fact, the minimum scale used in the 3$\times$2pt analysis is 6 $h^{-1}$Mpc, but due to their usage of analytical marginalization of small-scale information (see Section 4.2 in \citealt{y3-gglensing} and \citealt{MacCrann_2019}) the scales between 6-8 $h^{-1}$Mpc do not add significant information. Regardless, we use the large-scale ratios purely as validation for the small-scale ratios, as detailed in Section~\ref{sec:results}.

Small-scale ratios are defined using angular measurements below the minimum scale used in the cosmology analysis, i.e. 6 $h^{-1}$Mpc. Small-scale ratios will be our fiducial set of ratios, and next we focus on defining the lower boundary of scales to be used for those ratios. If ratios were purely geometrical, they would be scale-independent, and hence we could use all measured scales to constrain them (the full measured set of scales for galaxy-galaxy lensing in DES Y3 is described in \citealt{y3-gglensing}). However, as we saw in the previous Section, lensing ratios are not purely geometrical but there are other physical scale-dependent effects which need to be modeled accurately, and hence we are restricted to angular scales where the modeling is well characterized. In particular, in some ratio configurations there is enough overlap between lenses and sources to make the ratios sensitive to Intrinsic Alignments (IA), even if this dependence is smaller than for the full tangential shear measurement.

Figure \ref{fig:ratios_scale_IA} shows the impact of different IA models (different parameter choices for TATT) on all the lensing ratios considered. This Figure can be compared to Figure \ref{fig:ratios_scale}, and it is unsurprising to find that the impact of IA is smallest for the ratios that are closest to purely geometrical ratios. There are two cases in which the ratios are insensitive to IA (Figure \ref{fig:ratios_scale_IA}) and well modeled with geometry only (Figure \ref{fig:ratios_scale}), which correspond to the combinations ($l_1, s_3, s_4$) and ($l_2, s_3, s_4$). These ratios involve lens-source combinations with negligible overlap between lenses and sources and are thus not affected by IA. For these geometrical ratio combinations, we predict scale-independent ratios and hence we are able to accurately model the measurements at all scales, down to the minimum angular separation in which we measure the tangential shear, which is 2.5 arcmins \citep{y3-gglensing}.

For the remaining combinations we choose not to include physical scales below 2 $h^{-1}$Mpc, to avoid approaching the 1-halo regime. This decision is driven by the importance of shear ratio in constraining IA and the corresponding requirement to restrict the analysis to the range of validity of our fidicual IA model, the Tidal Alignment Tidal Torquing (TATT) model. This model captures nonlinear IA effects, notably the tidal torquing mechanism relevant for blue/spiral galaxies and the impact of weighting by source galaxy density, which becomes important on scales where clustering is non-negligible. The TATT model is thus significantly more flexible than the frequently-used nonlinear alignment model (NLA), which itself has been shown to accurately describe alignments of red galaxies down to a few Mpc (e.g.\ \citealp{Singh_2015, Blazek_2015}). However, as a perturbative description, the TATT model will not apply on fully nonlinear scales and thus is not considered robust within the 1-halo regime. While this choice of minimum scale is supported by both theoretical expectations and past observational results, we use our analysis when restricting to large scales as an additional robustness check. As shown in Figure~\ref{fig:ia_data}, the IA constraints from the large-scale shear ratio information are fully consistent with the fiducial shear ratio constraints, providing further support for our assumption that the TATT model can describe IA down the minimum scale. In Table~\ref{tab:scale_cuts} we summarize the scale cuts described in this section for each of the ratio combinations. Finally, it is worth noting that the choice of physical scales is the same for the two lens galaxy samples used in this work, but the choice of angular scales varies due to their slightly different redshift distributions.

\begin{table}
\centering
\setlength{\tabcolsep}{5pt}
\begin{tabular}{c|c|c|c}
\pmb{$(l_i,s_j,s_k)$}  &   \textbf{Scales} & $N_{dp}$ RM & $N_{dp}$ ML \\
\rule{0pt}{3ex} 
$(l_1,s_1,s_4)$ & 2 -- 6 $h^{-1}$Mpc &  4 & 5\\
$(l_1,s_2,s_4)$ & 2 -- 6 $h^{-1}$Mpc & 4 & 5\\
$(l_1,s_3,s_4)$ & 2.5 arcmin -- 6 $h^{-1}$Mpc & 10 & 10\\
\rule{0pt}{3ex} 
\rule{0pt}{3ex} 
$(l_2,s_1,s_4)$ & 2 -- 6 $h^{-1}$Mpc &  4 & 5\\
$(l_2,s_2,s_4)$ & 2 -- 6 $h^{-1}$Mpc & 4 & 5\\
$(l_2,s_3,s_4)$ &  2.5 arcmin -- 6 $h^{-1}$Mpc & 8 & 9\\
\rule{0pt}{3ex} 
\rule{0pt}{3ex} 
$(l_3,s_1,s_4)$ & 2 -- 6 $h^{-1}$Mpc &  4 & 5\\
$(l_3,s_2,s_4)$ & 2 -- 6 $h^{-1}$Mpc & 4 & 5\\ 
$(l_3,s_3,s_4)$ & 2 -- 6 $h^{-1}$Mpc & 4 & 5\\
\end{tabular}
\caption{Redshift bin combinations and scales used in this work for the ``small-scales'' lensing ratios probe. $(l_i,s_j,s_k)$ is a label that specifies the lens and source bins considered in the ratio, where $s_k$ is in the denominator. In general, we use scales between 2 -- 6 $h^{-1}$Mpc, except for the combinations which have almost no overlap between lenses and sources, for which we use all scales available (with a lower limit of 2.5 arcmins) since they are dominated by geometry and almost not affected by IA and magnification effects. $N_{dp}$ is the number of data points remaining after applying our scale cuts, which we show for both lens samples: \textsc{redMaGiC} (RM) and \textsc{MagLim} (ML) (the variations in $N_{dp}$ for both samples come from them having slightly different mean redshifts). \label{tab:scale_cuts}}
\end{table}

\subsubsection{Estimation of the ratio}
\label{sec:estimate}

Having defined the set of ratios of galaxy-galaxy lensing measurements to be used, and the set of angular scales to employ in each of them, now we will describe the procedure to measure the lensing shear ratios. Let $\gamma_t^{l_i,s_j}(\theta)$ and $\gamma_t^{l_i,s_k}(\theta)$ be two galaxy-galaxy lensing measurements as a function of angular scale ($\theta$) around the same lens bin $l_i$ but from two different source bins, $s_j$ and $s_k$, and we want to estimate the ratio of them. Since the ratios are mostly geometrical they are predominantly scale independent (Figure \ref{fig:ratios_scale}). We have checked that using ratios as a function of scale does not significantly improve our results (although that may change for future analyses with larger data sets). Therefore, for simplicity, we average over angular scales between our scale cuts in the following way:

\begin{equation}
\label{eq:ratio}
r^{(l_i,s_j,s_k)} \equiv \left \langle \frac{   \gamma_t^{l_i,s_j} (\theta)}{ \gamma_t^{l_i,s_k} (\theta) } \right \rangle_\theta =  \left \langle r^{(l_i,s_j,s_k)} (\theta) \right \rangle_\theta ,
\end{equation}

where the average over angular scales, $\left \langle ... \right \rangle_\theta$, includes the corresponding correlations between measurements at different angular scales.  We can denote the ratio measurements as a function of scale as vectors, such as $r^{(l_i,s_j,s_k)} (\theta) \equiv \mathbf{r}^{(l_i,s_j,s_k)}$, and $(l_i,s_j,s_k)$ is a label that specifies the lens and source bins considered in the ratio. In order to account for all correlations, we will assume we have a fiducial theoretical model for our lensing measurements, $\tilde{\gamma}_t^{l_i,s_j}(\theta)$ and $\tilde{\gamma}_t^{l_i,s_k}(\theta)$ and a joint covariance for the two measurements as a function of scale, $\mathbf{C}_{\tilde{\gamma}}$, such that

\begin{equation}
\left ( \mathbf{C}_{\tilde{\gamma}} \right )_{m,n}= \mathrm{Cov}[\tilde{\gamma}_t^{l_i,s_j}(\theta_m),\tilde{\gamma}_t^{l_i,s_k}(\theta_n)].
\end{equation}

Now we want to estimate the average ratio of lensing measurements. The ratio is a non-linear transformation, as it is clear from Equation (\ref{eq:ratio}). The covariance of the ratio as a function of scale can be estimated as
\begin{equation}
\mathbf{C}_{\mathbf{r}} =  \mathbf{J} \: \mathbf{C}_{\tilde{\gamma}} \: \mathbf{J}^T,   
\end{equation}
where $\mathbf{J}$ is the Jacobian of the ratio transformation as a function of scale from Equation (\ref{eq:ratio}), $\mathbf{r}^{(l_i,s_j,s_k)}$, and can be computed exactly using the theoretical model for the lensing measurements. Note that $\mathbf{C}_{\tilde{\gamma}}$, $\mathbf{C}_{\mathbf{r}}$ and $\mathbf{J}$ are all computed for a given ratio $(l_i,s_j,s_k)$. Having the covariance for the ratio as a function of scale, the estimate of the mean ratio having minimum variance is given by:

\begin{equation}
\label{eq:ratio_est}
\begin{aligned}
    r^{(l_i,s_j,s_k)} &= \sigma_r^2 \left ( \mathbf{D}^T \: \mathbf{C}^{-1}_{\mathbf{r}} \mathbf{r}^{(l_i,s_j,s_k)} \right ),\\
    \mathrm{with} \; \sigma_r^2 &= \left ( \mathbf{D}^T \: \mathbf{C}_{\mathbf{r}}^{-1 \: (l_i,s_j,s_k)} \: \mathbf{D} \right )^{-1}.
\end{aligned}
\end{equation}

Here $\mathbf{D}$ is a design matrix equal to a vector of ones, $[1,...,1]^T$, of the same length as $\mathbf{r}^{(l_i,s_j,s_k)}$ (the number of angular bins considered). Note that the estimator for the ratio in Equation (\ref{eq:ratio_est}) reduces to an inverse variance weighting of the angular bins for a diagonal covariance, and to an unweighted mean in the case of a diagonal covariance with constant diagonal values. 

\begin{figure}
 \centering
 \includegraphics[width=0.48\textwidth]{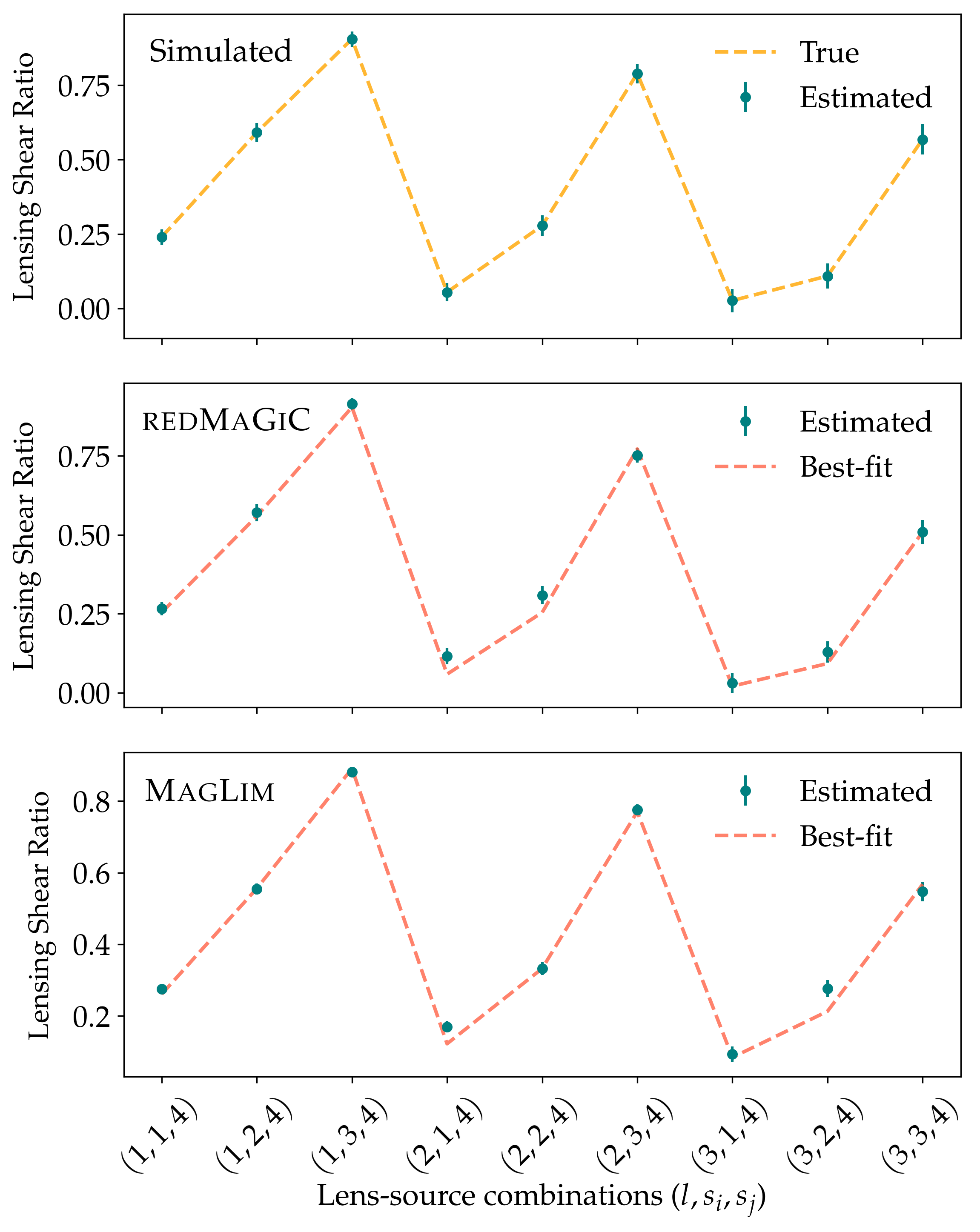}
 \caption{\emph{(Upper panel:)} True values of the ratios $\{ r \}$ for our fiducial theory model, together with the estimates of the simulated ratios using the measurement procedure described in \S\ref{sec:estimate} and the uncertainties estimated using the procedure described in \S\ref{sec:cov}. \emph{(Middle panel:)} Measured set of shear ratios and their uncertainties in the \textsc{redMaGiC} data, together with the best-fit model from the 3$\times$2 DES Y3 cosmological analysis of the \textsc{redMaGiC} sample ($\chi^2$/ndf = 11.3/9, $p$-value of 0.26). \emph{(Lower panel:)} Measured set of shear ratios and their uncertainties in the \textsc{MagLim} data, together with the best-fit model from the 3$\times$2 DES Y3 cosmological analysis of the \textsc{MagLim} sample ($\chi^2$/ndf = 18.8/9, $p$-value of 0.03, above the threshold for inconsistencies which we originally set at $p$-value = 0.01 for the DES Y3 analysis).  }
 \label{fig:ratios_sim_data}
 \end{figure}

For the fiducial simulated data in this work, we use the redshift distributions of the \textsc{redMaGiC} lens sample, although the differences between the \textsc{redMaGiC} and \textsc{MagLim} samples are small in the first three lens bins (see Figure \ref{fig:nzs}). Figure \ref{fig:ratios_sim_data} shows values of the fiducial estimated lensing shear ratios for both our simulated data and the real unblinded data. For the simulated case, we show the true values of the ratios, i.e.~those measured directly from the noiseless case, as well as the estimated ratios when noise is included. For the data cases, we show the fiducial set of data ratios used in this work for both \textsc{redMaGiC} and \textsc{MagLim} lens samples together with the corresponding best fit model using the full 3$\times$2 DES Y3 cosmological analyses. The data results will be discussed in detail in Section \ref{sec:results}. Next, we will describe the covariance estimate of the ratios and we will assess the performance of our estimator. 

\subsubsection{Covariance of the ratios}
\label{sec:cov}

We have described above how we compute the ratio of a given pair of galaxy-galaxy lensing measurements. Now, we describe how we compute the covariance between different ratios, from different pairs of lensing measurements. First, we use the fiducial model and theory joint covariance for all the galaxy-galaxy lensing measurements produced and validated in \citet{y3-covariances}, to produce $10^5$ covariant realizations of the galaxy-galaxy lensing measurements drawn from a mutivariate Gaussian centered at the fiducial model, and with the theoretical galaxy-galaxy lensing covariance. 

For each of these $10^5$ realizations of galaxy-galaxy lensing measurements, we measure the set of 9 shear ratios using the procedure described above. That yields $10^5$ realizations of the set of 9 ratios. We use that to compute the 9$\times$9 covariance of the ratios, which is shown in Fig.~\ref{fig:ratios_cov} in Appendix~\ref{sec:appendix_covariance}. The number of realizations we use to produce this covariance ($10^5$) is arbitrary, but we have checked that the results do not vary when using a larger number of realizations.

\subsection{Independence between small and large scales}\label{sec:independence_between_small_large}

In this section we discuss why the SR likelihood is independent of the 2pt likelihood. We also discuss the independence of the large-scale ratios defined in Section~\ref{sec:small-large-ratios} with respect to the small-scale ones. That independence will allow us to use the large-scale ratio information as validation of the information we get from small-scales.

The correlation of the SR likelihood with the (3$\times$)2pt likelihood will come mostly from the galaxy-galaxy lensing 2pt measurements. Since we do not leave any gap between the minimum scale used for 2pt measurements (6$h^{-1}$Mpc) and the small-scale ratios, this can in principle be worrying since the tangential shear is non-local, and therefore it receives contributions from physical scales in the galaxy-matter correlation function that are below the scale at which it is measured \citep{Baldauf2010, MacCrann_2019, park2020localizing}. However, for the large scales 2pt galaxy-galaxy lensing  used in the DES Y3 3$\times$2pt analysis, we follow the approach of \citet{MacCrann_2019} and marginalize analytically over the unknown enclosed mass, which effectively removes any correlation with scales smaller than the small-scale limit of 6 $h^{-1}$Mpc, ensuring that the information from the 3$\times$2pt measurements is independent from the small-scale ratios used in this work, which use scales smaller than 6 $h^{-1}$Mpc. We call this procedure ``point-mass marginalization''. This point-mass marginalization scheme significantly increases the uncertainties in galay-galaxy lensing measurements around 6-8 $h^{-1}$Mpc (see Figure 8 and Section 4.2 in \citet{y3-gglensing}). Also, see \citet{MacCrann_2019} and \citet{y3-2x2ptbiasmodelling} for a description of the point-mass marginalization implemented in the 3$\times$2pt analysis.

Regarding the large-scale shear ratios used in this work, we will only use scales larger than 8 $h^{-1}$Mpc, to ensure independence from the small-scale ratios using scales smaller than 6 $h^{-1}$Mpc (since for the SR-only chains we do not apply the point-mass marginalization, given that the ratios are not sensitive to first approximation to any enclosed mass). In order to assess the independence of small and large-scale ratios, we estimate the cross-covariance of the small and large-scale ratios, using again $10^5$ realizations of the galaxy-galaxy lensing measurements and deriving small and large-scale ratios for each of them, using the same procedure as in \$\ref{sec:cov}. We ensure that the corresponding $\Delta \chi^2$ due to including or ignoring the cross-covariance is smaller than 0.25. The reasons for this independence relate to the 2 $h^{-1}$Mpc gap left between small and large scales and to the importance of shape noise at these scales, which helps decorrelating different angular bins.

\subsection{Gaussianity of the SR likelihood}\label{sec:performance_estimator}

\begin{figure}
 \centering
 \includegraphics[width=0.5\textwidth]{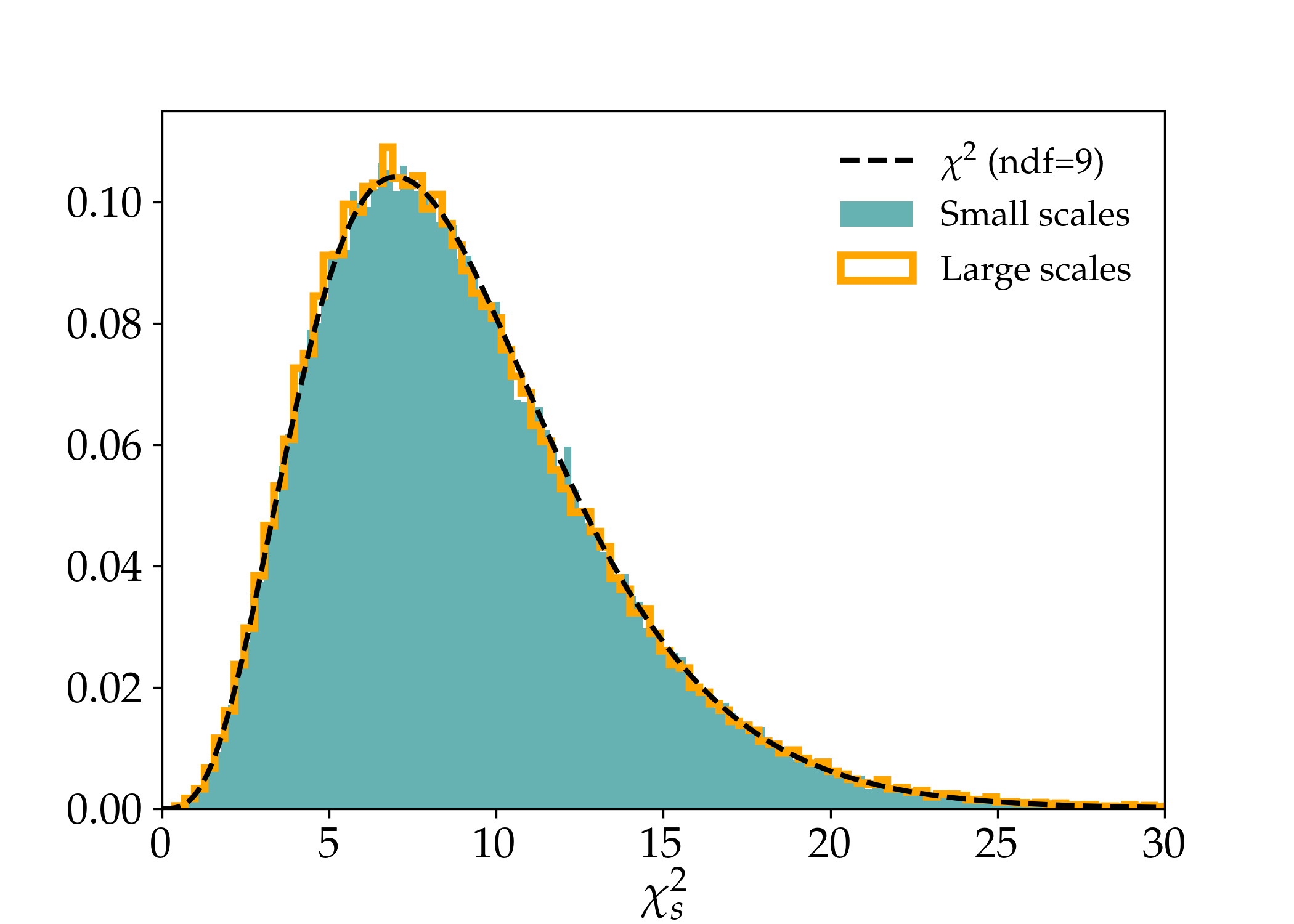}
 \caption{Distribution of $\chi^2_s$ from Equation (\ref{eq:chi2s}), for small and large-scale ratios, compared to a chi squared distribution with a number of degrees of freedom equal to the numbers of ratios in $\{ r \}_s$. }
 \label{fig:chi2s}
 \end{figure}

Because the shear ratios are a non-linear transformation of the galaxy-galaxy lensing measurements, it is important to test the assumption of Gaussianity in the likelihood.  We have a number of realizations $s$ of the lensing measurements, drawn from the theory curves and the corresponding covariance, and for each of them we have a set of 9 measured ratios, $\{ r \}_s$, and we have a 9$\times$9 covariance for those ratios, which we can denote $\mathbf{C}_{\{ r \}}$. Importantly, we also have the set of ratios for the noiseless fiducial model used to generate the realizations, denoted by $\{ r \}_0$. Figure \ref{fig:ratios_sim_data} shows the noiseless (or true) ratios from the model, $\{ r \}_0$, together with the estimated mean and standard deviation of the noisy ratios, $\{ r \}_s$. Also, if the likelihood of observing a given set of noisy ratios given a model, $p(\{ r \}_s|\{ r \}_0)$, is Gaussian and given by the covariance $\mathbf{C}_{\{ r \}}$, then the following quantity:
\begin{equation}
\label{eq:chi2s}
    \chi^2_s = (\{ r \}_s-\{ r \}_0) \: \mathbf{C}_{\{ r \}}^{-1} \: (\{ r \}_s-\{ r \}_0)^T
\end{equation}
should be distributed like a chi squared distribution with a number of degrees of freedom equal to the number of ratios (9 in this case), so $\chi^2_s \sim \chi^2(x,\mathrm{ndf}=9)$. Figure \ref{fig:chi2s} shows the agreement between the distribution of $\chi_s$ compared to the expected chi squared distribution for ratios at small and large scales. This agreement demonstrates that our likelihood for the ratios is Gaussian, and in conjunction with Figure \ref{fig:ratios_sim_data}, it provides validation that our estimator does not suffer from any significant form of noise bias.

 \section{Validation of the model}
 \label{sec:model_validation}

In this section we validate our model for the lensing ratios by exploring the impact of several effects that are not included in our fiducial model, which are relevant to galaxy-galaxy lensing measurements at small scales (the corresponding validation for other DES Y3 data vectors is performed in \citealt{y3-generalmethods}). The fiducial model is described in \S\ref{sec:model}. The effects we consider are in some cases explored directly at the theory level (e.g.~by changing the input power spectrum) or using ratios measured in realistic $N$-body simulations. In this way, all the tests in this section are performed using noiseless simulated data vectors except for the Buzzard case (in \S\ref{sec:validation_buzzard}) which includes noise. For testing purposes, we will analyze the impact of such effects in the ratios, at both small and large scales, but we will also assess their impact on the derived constraints on our model parameters using the same priors we use in the data, which are the most relevant metric. For that, we will perform MCMC runs as described in \S\ref{sec:parameter_inference_methodology} for the various effects under consideration. The priors and allowed ranges of all parameters in our model are described in Table \ref{tab:model_validation_priors} and aim to mimic the configuration used for the final runs in the data, described in \S\ref{sec:results}. A summary of the resulting constraints for each test is included in Figures \ref{fig:summ1} and \ref{fig:ia_model_validation}, while further details are included in each subsection. 

\subsection{Fiducial simulated constraints}

For the purpose of comparison and reference, in the figures of this section we also include constraints from the fiducial SR case, where ratios are constructed directly using the input theory model. As we did in the previous section, we use the redshift distributions of the \textsc{redMaGiC} lens sample for the fiducial simulated ratios, although the differences between the \textsc{redMaGiC} and \textsc{MagLim} samples are small in the first three lens bins (see Figure \ref{fig:nzs}). We have generated our fiducial simulated data vector using the best-fit values of the 3$\times$2pt+SR results for the cosmological parameters and the IA and galaxy bias parameters\footnote{Specifically, we use the values of the first 3$\times$2pt+SR \textsc{redMaGiC} unblinded results and the halo-model covariance evaluated at these values.}. 

In addition, the fiducial case allows us to determine what parameters are being constrained using the information coming from the ratios. Figure \ref{fig:summ1} shows the constraints on the parameters corresponding to source redshifts, source multiplicative shear biases and lens redshifts. Due to the strong priors imposed on these parameters, no correlations are observed between them and hence we show the marginalized 1-D posteriors. In the fiducial SR case, the ratios improve the constraints on the parameters corresponding to source redshifts, while shear calibration and lens redshifts are not significantly constrained beyond the priors imposed on those parameters (Table~\ref{tab:model_validation_priors}). In detail, for the four source redshift parameters in Figure \ref{fig:summ1}, the posteriors using the ratios improve the prior constraints on $\Delta z_s$ by 12\%, 25\%, 19\% and 8\% respectively for each bin. Furthermore, the ratios are able to place constraints on some of the intrinsic alignments (IA) parameters of our model, for which we do not place Gaussian priors. For IA, out of the 5 parameters in the model, the ratios are most effective at constraining a degeneracy direction between IA parameters $a_1$ and $a_2$, as in Figure \ref{fig:ia_model_validation} (see \citealt{y3-gglensing} or \citealt*{y3-cosmicshear2} for a full description of the IA model used or Section~\ref{sec:model} in this paper for a summary of the most relevant equations). These IA constraints from SR will become important for constraining cosmological parameters when SR is combined with other probes like cosmic shear (see Section \ref{sec:combination}).

\subsubsection{Large-scale ratios (LS)}

\begin{figure*}
 \centering
 \includegraphics[width=0.95\textwidth]{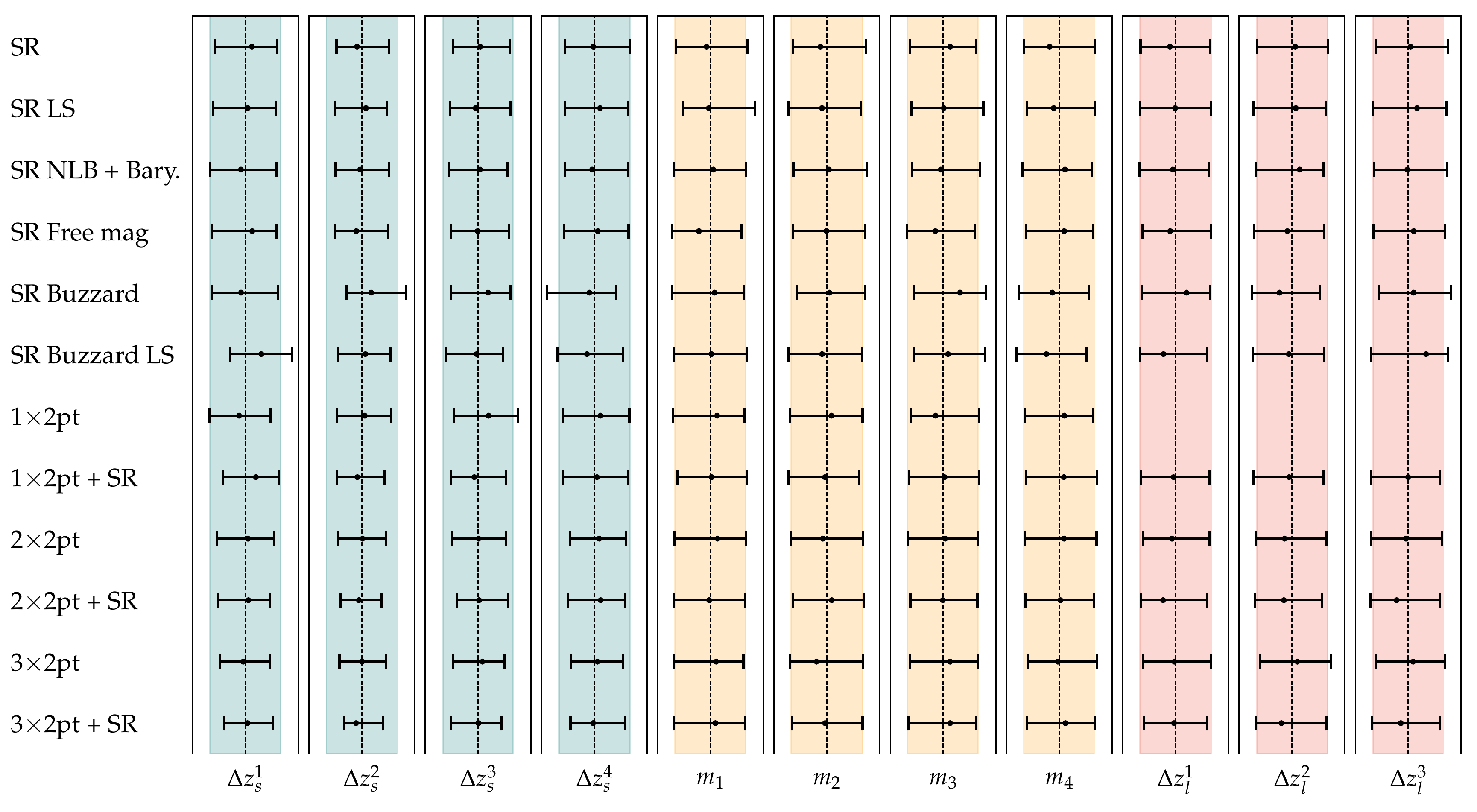}
 \caption{Summary of the posteriors on the model parameters corresponding to source redshifts, shear calibration and lens redshifts for different SR-only test runs described in Section \ref{sec:model_validation} and combination runs from Section~\ref{sec:combination}. All the above tests are performed using noiseless simulated data vectors except for the Buzzard ones which include noise. The coloured bands show the 1$\sigma$ prior in each parameter, while the black errorbars show 1$\sigma$ posteriors. }
 \label{fig:summ1}
 \end{figure*}

\begin{figure}
 \centering
 \includegraphics[width=0.45\textwidth]{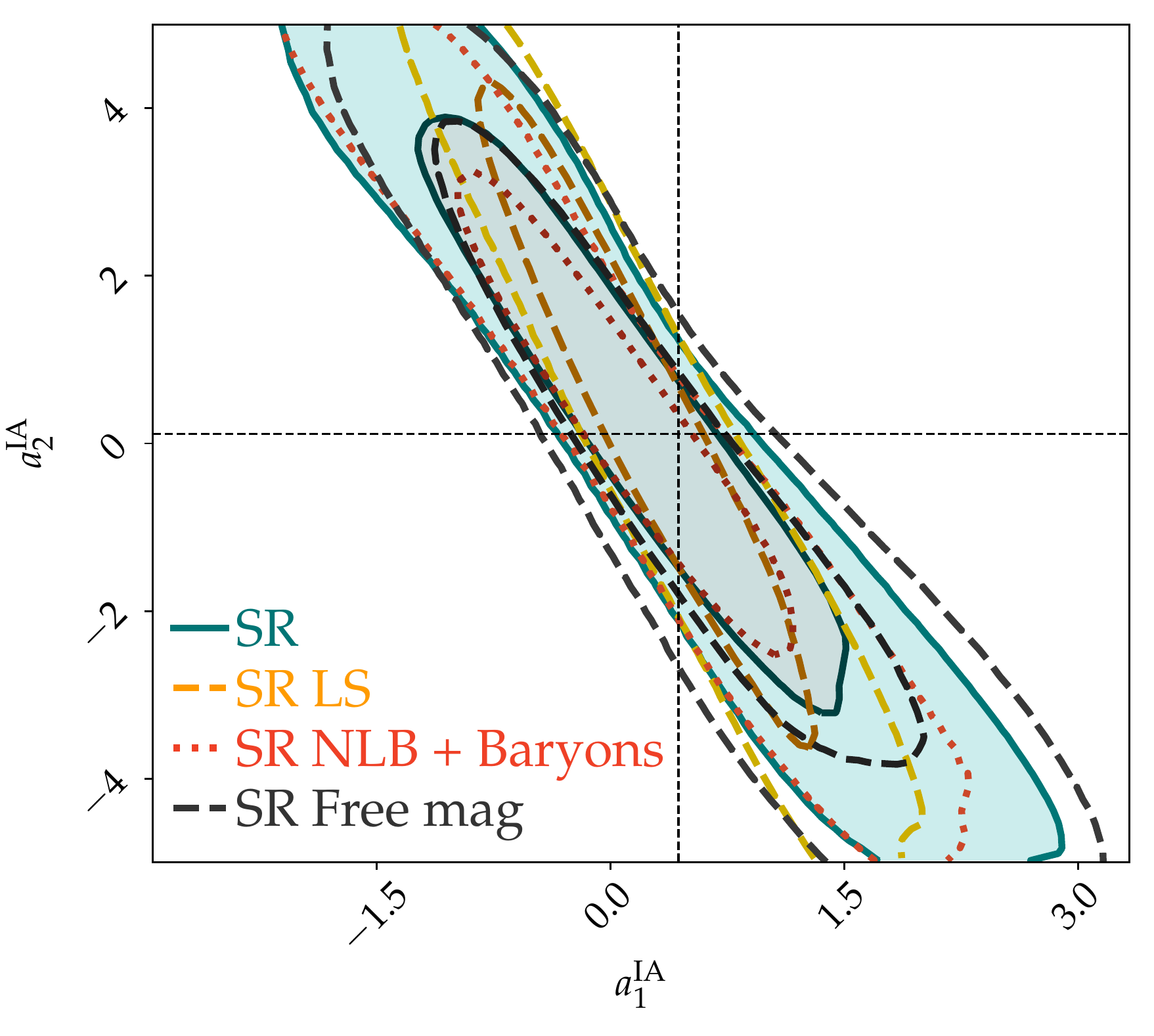}
 \caption{This plot summarizes the posteriors on the two intrinsic alignment model parameters that are constrained by the ratios, for different SR only test runs described in Section \ref{sec:model_validation}, using noiseless simulated data vectors. }
 \label{fig:ia_model_validation}
 \end{figure}

One important test that will be used as a direct model validation in the data is the comparison of model posteriors using ratios from small and large scales. This comparison is interesting because the model for galaxy-galaxy lensing is more robust at large scales, and because small and large scales are uncorrelated since they are dominated by shape noise. Therefore, we can compare our fiducial model constraints coming from small-scale ratios to the corresponding constraints from large scales, and a mismatch between these will point out a potential problem in the modeling of small scales. In Figures \ref{fig:summ1} and \ref{fig:ia_model_validation}, we show the model constraints from  large-scale ratios, for reference. We will also perform this test for the results on $N$-body simulations in Section~\ref{sec:validation_buzzard}, and directly on the data in Section~\ref{sec:results:SRonly}.

\begin{table*}

\centering
\addtolength{\tabcolsep}{10pt} 
\begin{tabular}{l|c|c}
                     &         \textbf{Range}             &        \textbf{Prior}             \\
\rule{0pt}{3ex}Source redshifts $\Delta z_s^j$                     &         [-0.1,0.1]             &        $\mathcal{N}$(0, [0.018,0.015,0.011,0.017])              \\
Shear calibration $m^j$                     &         [-0.1,0.1]             &       $\mathcal{N}$(0, [0.0091,0.0078,0.0076,0.0076])               \\
Lens redshifts $\Delta z_l^i$                     &         [-0.05,0.05]             &      $\mathcal{N}$(0, [0.004,0.003,0.003])                \\
Galaxy bias $b^i$                     &         [0.8,3.0]             &     Uniform                \\
IA $a_1, a_2, \alpha_1, \alpha_2$                     &         [-5,5]             &     Uniform                 \\
IA bias TA                     &         [0,2]             &     Uniform                 
\end{tabular}
\addtolength{\tabcolsep}{-10pt} 
\caption{Allowed ranges and priors of the model parameters for the chains run in Sections \ref{sec:model_validation} and \ref{sec:combination}. Indices $i$ in the labels refer to the 3 lens redshift bins, and indices $j$ refer to the 4 source redshift bins, all defined in Section \ref{sec:data}.   } \label{tab:model_validation_priors}
\end{table*}

\subsection{Baryons and non-linear galaxy bias}\label{sec:baryons_and_nlbias}
Hydrodynamical simulations suggest that baryonic effects, specifically the ejection of gas due to feedback energy from active galactic nuclei (AGN), have an impact on the matter distribution at cosmologically relevant scales \citep{Mead_2015}. Such effects may lead to differences in the galaxy-galaxy lensing observable at the small scales considered in this work.  In order to test this effect, we model it rescaling the non-linear matter power spectrum with the baryonic contamination from OWLS (OverWhelmingly Large Simulations project, \citealt{OWLS, vanDaalen11}) as a function of redshift and scale. Specifically, to obtain the baryonic contamination, we compare the power spectrum from the dark matter-only simulation with the power spectrum from the OWLS AGN simulation, following \citet{y3-generalmethods}.

In addition, non-linear galaxy bias effects would potentially produce differences in the ratios that could be unexplained by our fiducial model. We utilize a model for non-linear galaxy bias that has been calibrated using $N$-body simulations and is described in \citet{Pandey_2020,y3-2x2ptbiasmodelling}. In order to test the impact of these effects on the ratios, we produce a set of simulated galaxy-galaxy lensing data vectors including the effects of baryons and non-linear galaxy bias as described above, and produce the corresponding set of shear ratios. Overall we use the same procedure which is used in \citet{y3-generalmethods} to contaminate the fiducial data vector with these effects and propagate this contamination to the ratios. Then, we derive constraints on our model parameters using this new set of ratios, and we show the results in Figures \ref{fig:summ1} and \ref{fig:ia_model_validation}. In those figures we can see the small impact of these effects in our constraints compared to the fiducial case, confirming that baryonic effects and non-linear galaxy bias do not significantly bias our model constraints from the ratios.

\subsection{Halo Occupation Distribution Model} \label{sec:hod}

The Halo Occupation Distribution (HOD, \citealt{COORAY2002}) model provides a principled way of describing the connection between galaxies and their host dark matter halos, and it is capable of describing small-scale galaxy-galaxy lensing measurements at a higher accuracy than the Halofit approach, which is used in our fiducial model for the shear ratios (described in \S\ref{sec:model}). Next we test the differences between HOD and Halofit in the modeling of the ratios, and assess their importance compared to the uncertainties we characterized in Section \ref{sec:measurement}. We aim at performing two tests to assess the robustness of the fiducial Halofit modeling by comparing it to two HOD scenarios. One scenario showing the effect of HOD modeling, with a fixed HOD for each lens redshift bin, and another including HOD evolution within each lens redshift bin. For these tests, we perform the comparisons to a fiducial model without intrinsic alignments and lens magnification, for simplicity and to isolate the effects of HOD modeling compared to Halofit. 

For these tests we use the MICE $N$-body simulation where a DES-like lightcone catalog of \textsc{redMaGiC} galaxies with the spatial depth variations matching DES Y3 data is generated (see Section \ref{sec:mice}). Using this catalog, we measure the mean HOD of the galaxies in the five redshift bins (\S\ref{sec:data}) as well as in higher resolution redshift bins with $\delta z \sim 0.02$. Note that these measurements are done using true redshifts of the galaxies, in order to pick up the true redshift evolution. We use these two measurements to predict the galaxy-galaxy lensing signal using a halo model formalism as described in the Appendix~\ref{app:HOD}. 

The upper panel of Figure \ref{fig:hod} shows the difference between the simulated ratios using the fiducial model and the simulated ratios obtained using a mean HOD model for each redshift bin, for both small and large scales. As expected, the difference for large scales is negligible ($\Delta \chi^2$ = 0.02), since the fiducial Halofit modeling is known to provide an accurate description of galaxy-galaxy lensing at large scales ($>8$ Mpc/$h$). At small scales (between 2 and 6 Mpc/$h$), we see very small deviations of the HOD simulated ratios compared to the fiducial ones ($\Delta \chi^2$ = 0.22 for 9 data points), which do not significantly alter the constraints on the model parameters when using the HOD-derived ratios. It is worth noting that this is not a trivial test since the effect on the tangential shear itself is very significant on these scales, as can be seen in figure~8 from \citet{y3-gglensing}.

The lower panel of Figure \ref{fig:hod} shows the difference of shear ratios produced by using a mean HOD for each redshift bin and an evolving HOD as obtained by high resolution measurements in MICE. We find a residual $\Delta \chi^2$ of 0.04 at small scales (even smaller at large scales) and hence consistent shear ratio estimates. Given the results shown in Figure \ref{fig:hod}, we conclude that non-linearities introduced by HOD evolution within a tomographic redshift bin will not bias our shear ratio estimates. 

It is important to note that the HOD tests described in this section correspond to one of our two lens samples, the \textsc{redMaGiC} sample, and that we do not show the equivalent test for the \textsc{MagLim} lens sample. However, having validated this for one of the lens samples, we will test the consistency between the SR constraints obtained with the two lens samples in Section \ref{sec:results}, and also \citet*{y3-cosmicshear1}  performs the same validation test in the combination of SR with cosmic shear.

\begin{figure}
 \centering
 \includegraphics[width=0.47\textwidth]{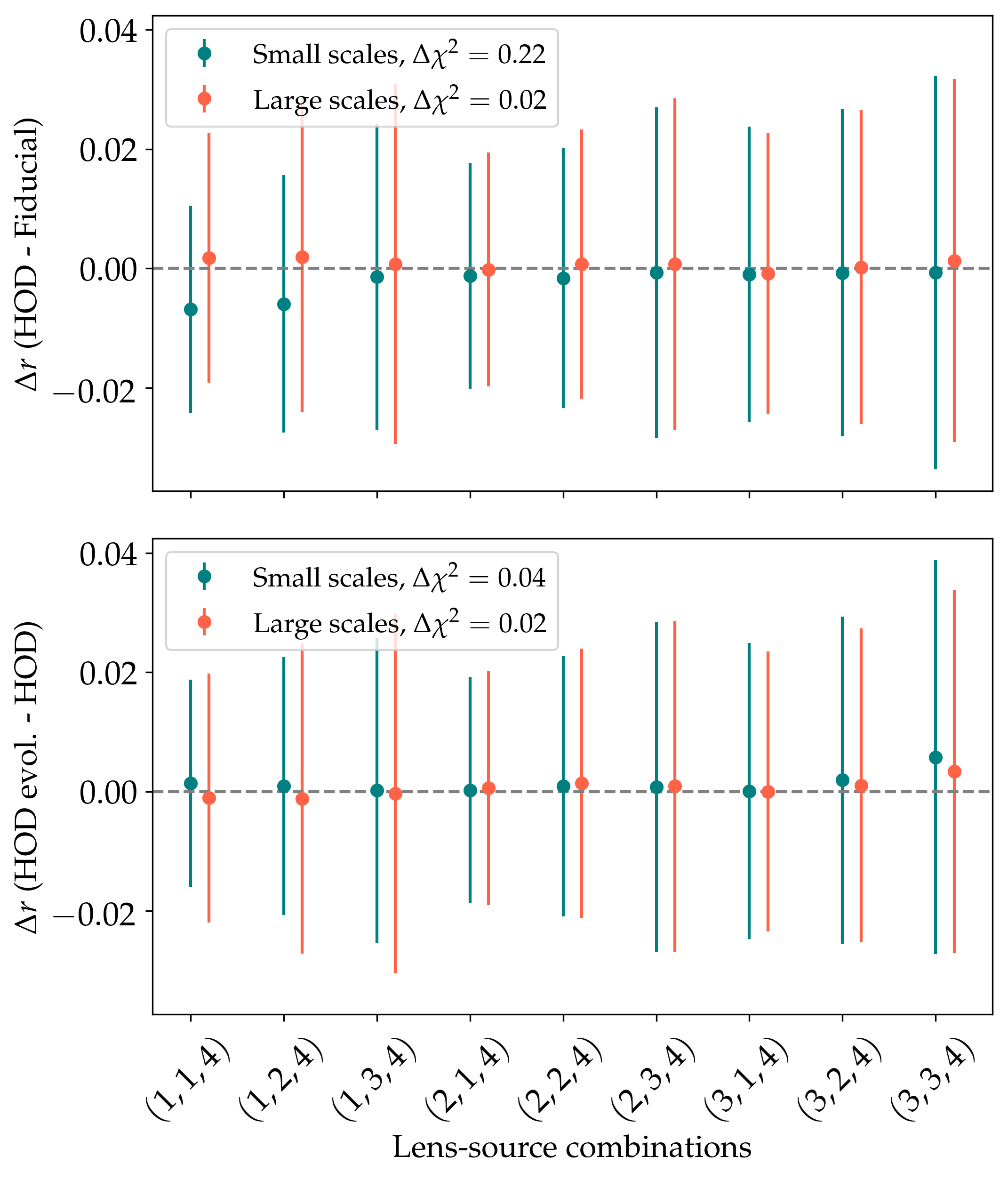}
 \caption{Effects of HOD modeling and HOD evolution on the shear ratios, for both small and large angular scales. The error bars show the ratio uncertainties from the same covariance as used in the data. }
 \label{fig:hod}
 \end{figure}

\subsection{Lens magnification}

The theoretical modeling of the galaxy-galaxy lensing signal and hence of the lensing ratios used in this work includes the effects of lens magnification. In the fiducial case, the lens magnification coefficients are fixed to the ones estimated using the \textsc{Balrog} software \citep{y3-balrog} in \citet{y3-2x2ptmagnification}. Here, we test the effect of letting the lens magnification coefficients be free for the SR analysis. In particular, Figure \ref{fig:summ1} and \ref{fig:ia_model_validation} test the effects of that choice (labelled there as ``Free mag'') on the parameters corresponding to lens and source redshifts, shear calibration and intrinsic alignments. No significant biases are observed and the derived constraints are comparable to the constraints using fixed lens magnification coefficients.

\subsection{Cosmology dependence}\label{sec:cosmology_dependence}

The lensing ratios themselves have very little sensitivity to cosmology. If they help with cosmological inference, it is because they help constrain some of the nuisance parameters that limit the cosmological constraining power. Here we will show their exact dependency, and that it is indeed safe to fix cosmological parameters when running SR only chains. Despite this weak dependency, the cosmological parameters are set as free parameters when the SR likelihood is run together with the 2pt likelihood (when combined with cosmic shear and the other 2pt functions). Hence even the small sensitivity of the ratios to cosmology is properly handled in the runs together with the 2pt likelihoods.  

In Fig.~\ref{fig:red_boosts_cosmo} we show how the lensing ratios change as a function of $\Omega_m$. Our fiducial simulated data vector assumes $\Omega_m \simeq 0.35$ and we show that varying that to $\Omega_m = 0.30$ or to $\Omega_m = 0.40$ has very little impact on the ratios, compared with their uncertainties, yielding $\Delta \chi^2 = 0.03, 0.01$, respectively, for 9 data points. We have also tested the dependency on the $\sigma_8$ parameter and found the lensing ratios change less than 0.5\% when changing from $\sigma_8=0.7$ to $\sigma_8=0.8$. 

\subsection{Boost factors and IA}\label{sec:boosts}

Boost factors are the measurement correction needed to account for the impact of lens-source clustering on the redshift distributions. When there is lens-source clustering, lenses and sources tend to be closer in redshift than represented by the mean survey redshift distributions that are an input to our model. This effect is scale dependent, being larger at small scales where the clustering is also larger. See Eq.~(4) of \citet{y3-gglensing} for its definition, related to the tangential shear estimator. 

We include boost factors as part of our fiducial measurements as detailed in \citet{y3-gglensing} (see their figure~3 for a plot showing the boost factors). However, since boost factors are more sensitive to some effects which are not included in our modeling, such as source magnification, it is useful to test their impact on the ratios. In Fig.~\ref{fig:red_boosts_cosmo} we show the difference in the ratios when including or not the boost factor correction and find it has a small impact on the ratios compared with their uncertainty, with $\Delta \chi^2 = 0.16$. 

Checking the influence of the boost factors on the ratios is also giving us an order-of-magnitude estimate of how much the nonlinear source clustering is impacting the signal, in particular in relation to intrinsic alignments. The IA term receives contributions from both the alignment of galaxies and the fact that sources cluster around lenses, leading to an excess number of lens-source pairs. We account for this term using the $b_{\text{TA}}$ parameter of the TATT model but on the smallest scales, roughly below a few Mpc, the TATT model will not sufficiently capture the non-linear clustering and IA \citep{Blazek_2015}. The fact that boost factors do not have a large impact on the ratios gives us an indication that our fiducial TATT model will suffice over the scales we use to construct the ratios.  

\subsection{Higher-order lensing effects}\label{sec:higher_order}

\begin{figure}
 \centering
 \includegraphics[width=0.48\textwidth]{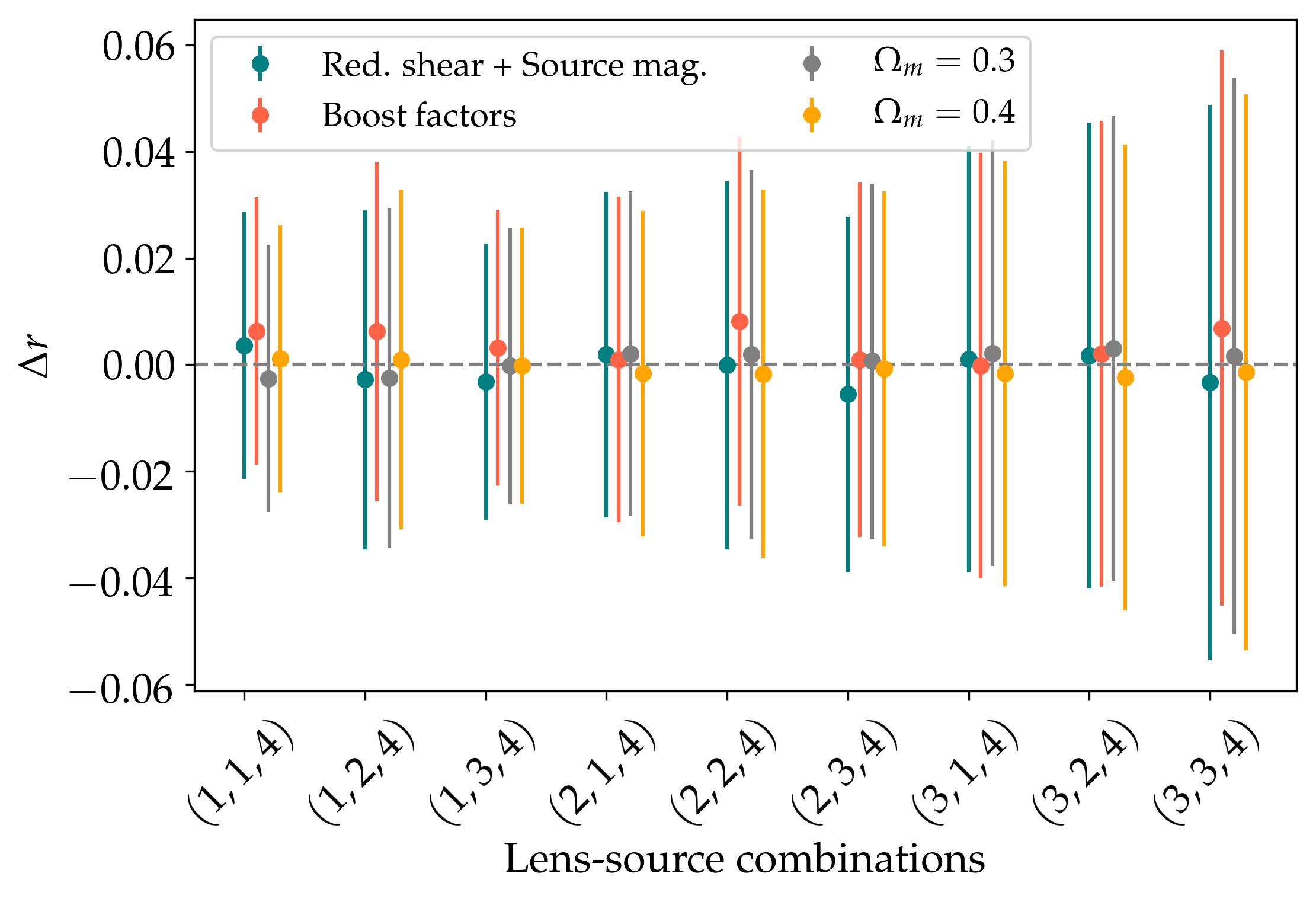}
 \caption{Impact of different effects on the lensing ratios, including cosmology dependence (see Sec.~\ref{sec:cosmology_dependence}), boost factors (see Sec.~\ref{sec:boosts}) and reduced shear + source magnification (see Sec.~\ref{sec:higher_order}). All these tests use noiseless simulated data vectors, and the error bars show the ratio uncertainties from the same covariance as used in the data. }
 \label{fig:red_boosts_cosmo}
 \end{figure}

In this section we test the impact of higher-order lensing effects to our model of the ratios, such as using the reduced shear approximation and not including source magnification in our model. In order to do that, we will propagate to the ratios the model developed and described in detail in \citet{y3-generalmethods} to include the combination of reduced shear and source magnification effects. This model is computed with the \textsc{CosmoLike} library \citep{Krause_2017} using a tree-level bispectrum that in turn is based on the non-linear power spectrum. For the source magnification coefficients, we use the values computed in \citet{y3-2x2ptmagnification}. In \citet{y3-gglensing}, the reduced shear contamination is illustrated for the tangential shear part. Here, we propagate that model to the lensing ratios, showing the small differences they produce on the ratios in Fig.~\ref{fig:red_boosts_cosmo}, with $\Delta \chi^2 = 0.09$ for 9 data points. The reduced shear contamination only produces a $\Delta \chi^2 = 0.02$ and therefore most of the change is coming from the source magnification part.

\subsection{Validation using $N$-body sims}\label{sec:validation_buzzard}

In this section we have so far considered different physical effects and tested their impact on the ratios at the theory level, for instance changing the input power spectrum used to generate the galaxy-galaxy lensing estimates. Now, instead, we use the Buzzard realistic $N$-body simulations, described in \S\ref{sec:buzzard}, to measure the lensing signal and the ratios which we then analyze using the fiducial model. These simulations are created to mimic the real DES data and hence they implicitly contain several physical effects that could potentially affect the ratios (e.g.~non-linear galaxy bias or redshift evolution of lens properties). For that reason, they constitute a stringent test on the robustness of our model. In addition, the tests in this part will be subject to noise in the measurement of the lensing signal and the ratios, due to shot noise and shape noise in the lensing sample, as opposed to the tests above which were performed with noiseless theoretical ratios. That measurement noise will also propagate into noisier parameter posteriors. 

In Figure \ref{fig:summ1} we include the results of the tests using $N$-body simulations, named SR Buzzard for the fiducial small-scale ratios and SR Buzzard LS for the large-scale SR test. The results are in line with the other tests in this Section, showing the robustness of the SR constraints also on $N$-body simulations (considering the fact that the Buzzard constraints include noise in the measurements, as stated above). In addition, due to the fact that there are no intrinsic alignments in Buzzard, and the fact that lens magnification is not known precisely, we do not show IA or magnification constraints from the Buzzard run.

\section{Combination with other probes and effect on cosmological constraints}
\label{sec:combination}

In the previous section we explored the constraining power of the lensing ratios defined in this work and we validated their usage by demonstrating their robustness against several effects in their modeling. However, in the DES Y3 cosmological analysis, lensing ratios will be used in combination with other probes. For photometric galaxy surveys, the main large-scale structure and weak lensing observables at the two-point level are
galaxy clustering (galaxy-galaxy), galaxy-galaxy lensing (galaxy-shear) and cosmic shear
(shear-shear), which combined are referred to as 3$\times$2pt. 
In this section, we will explore the constraining power of ratios when combined with such probes in DES, with the galaxy-galaxy lensing probe using larger scales compared to the lensing ratios.  

When used by themselves, lensing ratios have no significant constraining power on cosmological parameters, however, when combined with other probes, they can help constrain cosmology through the constraints they provide on nuisance parameters such as source mean redshifts or intrinsic alignments (IA). Next we will show simulated results on the impact of the addition of SR to the three 2pt functions used in the DES Y3 cosmological analysis. We will analyze the improvement in the different nuisance parameters but also directly on cosmological parameters. 

\subsection{SR impact on cosmic shear}

Cosmic shear, or simply 1$\times$2pt, measures the correlated distortion in the shapes of distant galaxies due to gravitational lensing by the large-scale structure in the Universe. It is sensitive to both the growth rate and the expansion history of the Universe, and independent of galaxy bias. Here we explore the constraining power of DES Y3 cosmic shear in combination with SR using simulated data. For that, we run MCMC chains where we explore cosmological parameters and the nuisance parameters corresponding to source galaxies, such as intrinsic alignments, source redshift calibration and multiplicative shear biases. Also, when using SR in combination with 1$\times$2pt, we sample over lens redshift calibration and galaxy bias parameters for the three redshift bins included when building the ratios, even if the posteriors on the galaxy bias are unconstrained. We make this choice to be fully consistent with the tests we have performed in the previous section but the results are consistent if we fix the galaxy bias parameters. For the lens redshift calibration parameters, we use the same priors detailed in the previous section. 

\begin{figure}
 \centering
 \includegraphics[width=0.48\textwidth]{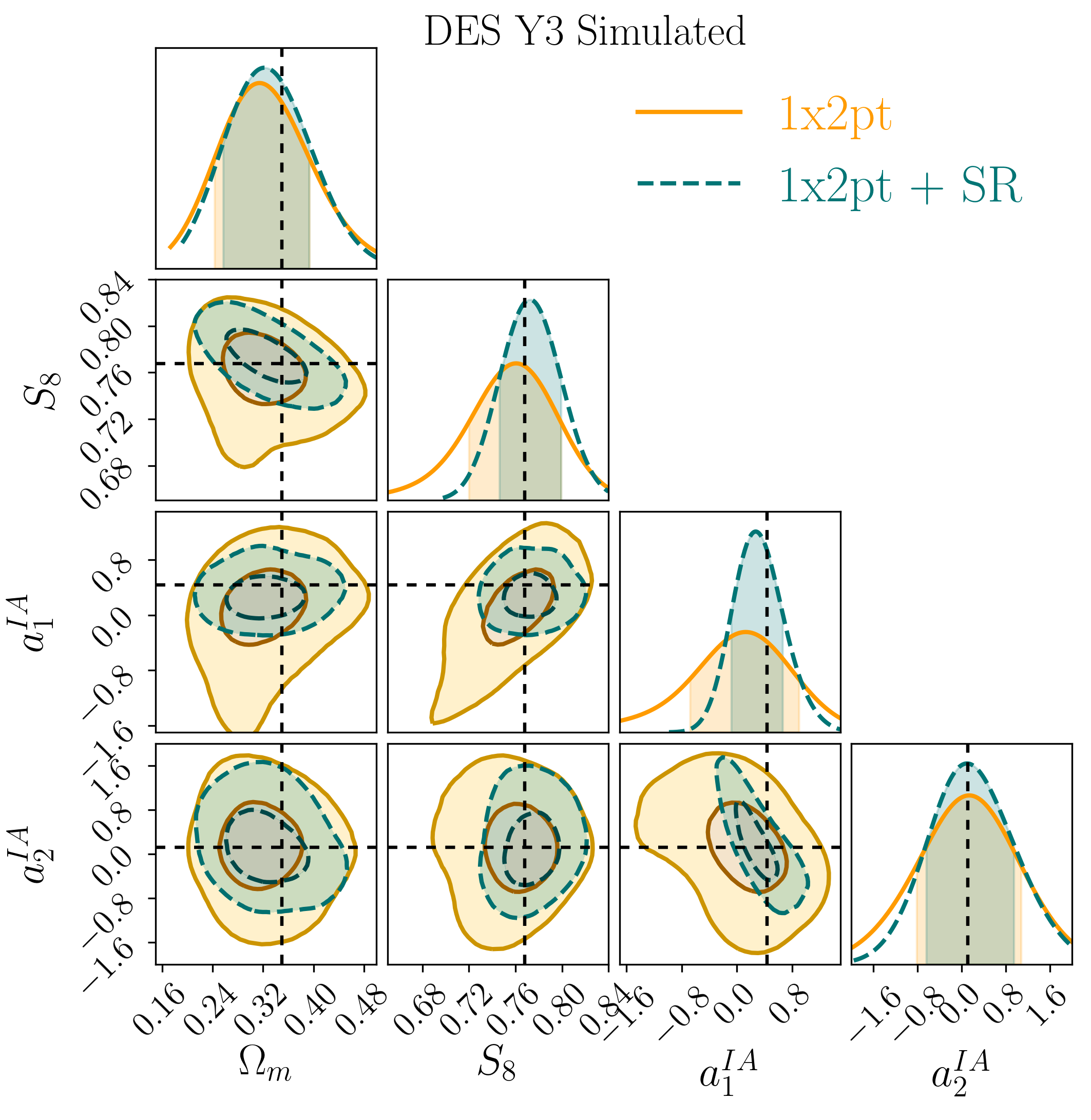}
 \caption{Simulated likelihood analysis showing the constraints on cosmological parameters $S_8$ and $\Omega_m$ and intrinsic alignments parameters $a_1^{IA}$ and $a_2^{IA}$ from cosmic shear only (1$\times$2pt) and cosmic shear and lensing ratios (1$\times$2pt + SR). }
 \label{fig:1x2sim}
 \end{figure}

The effect of adding SR to 1$\times$2pt is shown in Figure \ref{fig:1x2sim} for cosmological and IA parameters and in Figure \ref{fig:summ1} for the other nuisance parameters. For source redshift parameters, SR improves the constraints of all four source bins by 9\%, 13\%, 14\% and 2\%, respectively. Most importantly, SR significantly helps improve the constraints on cosmology, by about $25\%$ on $S_8$ and $3\%$ on $\Omega_m$ (see Table \ref{tab:cosmo_sim}). From Figure \ref{fig:1x2sim}, it is apparent the improvement in cosmology comes mostly from a major improvement in constraining the amplitudes of the IA modeling. The effect on the other IA parameters is shown in Appendix~\ref{sec:app_ia}.

\begin{table}
    \centering
    \label{tab:cosmo_sim}
\setlength{\tabcolsep}{5pt}
	\renewcommand{\arraystretch}{1.5}
    \begin{tabular}{ccc}
        \hline
		 & $\Delta \Omega_m$ & $\Delta S_8$ \\ 
		 \hline
1$\times$2pt & $-0.057^{+0.077}_{-0.038}$ & $-0.005^{+0.026}_{-0.030}$ \\ 
		1$\times$2pt + SR & $-0.050^{+0.078}_{-0.034}$ & $0.002^{+0.024}_{-0.018}$ \\ 
		2$\times$2pt & $0.008^{+0.028}_{-0.046}$ & $-0.019^{+0.044}_{-0.027}$ \\ 
		2$\times$2pt + SR & $-0.002^{+0.035}_{-0.037}$ & $-0.006^{+0.031}_{-0.037}$ \\ 
		3$\times$2pt & $-0.006^{+0.038}_{-0.021}$ & $0.003^{+0.013}_{-0.022}$ \\ 
		3$\times$2pt + SR & $0.011^{+0.018}_{-0.038}$ & $-0.006^{+0.021}_{-0.013}$ \\ 
		 \hline
    \end{tabular}
    \caption{Impact of SR on cosmological constraints using simulated DES Y3 data. The table shows parameter differences with respect to the truth values for the simulated data, which are $\Omega_m = 0.350$ and $S_8 = 0.768$. }
\end{table}

\subsection{SR impact on galaxy clustering and galaxy-galaxy lensing}

The combination of galaxy clustering and galaxy-galaxy lensing, also named 2$\times$2pt, is a powerful observable as it breaks the degeneracies between cosmological parameters and galaxy bias. When using SR in combination with 2$\times$2pt, there is no need to sample over additional parameters, and we use the same priors detailed in the previous section. 

When we add SR to the DES Y3 2$\times$2pt combination there is a modest improvement in constraining power for cosmological parameters, by about $4\%$ on $S_8$ and $3\%$ on $\Omega_m$ (see Table \ref{tab:cosmo_sim}). The reason this improvement is smaller than for the cosmic shear case is due to the fact that the 2pt galaxy-galaxy lensing measurements are already providing IA information in this case. This makes the change in the IA parameters when we add SR smaller, as shown in Appendix~\ref{sec:app_ia}. In Figure \ref{fig:summ1} we show the impact of adding SR for the other nuisance parameters. For source redshift parameters, SR improves the constraints of the first three source bins by 9\%, 14\% and 4\%, respectively. There is also a modest improvement on the other nuisance parameters as shown in Figure \ref{fig:summ1}.

\subsection{SR impact on 3$\times$2pt}

A powerful and robust way to extract cosmological information from imaging galaxy surveys involves the full combination of the three two-point functions, in what is now the standard in the field, and referred to as a 3$\times$2pt analysis. This combination helps constraining systematic effects that influence each probe differently. When using SR in combination with 3$\times$2pt, there is no need to sample over additional parameters, and we use the same priors detailed in the previous section. 

The effect of adding SR to the DES Y3 3$\times$2pt analysis is similar as for the 2$\times$2pt case. For cosmological parameters, there is an improvement in constraining power of about $3\%$ on $S_8$ and $5\%$ on $\Omega_m$ (see Table \ref{tab:cosmo_sim}). In Figure \ref{fig:summ1} we show the impact for the other nuisance parameters. For example, for source redshift parameters, SR improves the constraints of the second source bin by more than 15\%. The effect on the IA parameters is shown in Appendix~\ref{sec:app_ia}.

\section{Results with the DES Y3 data}
\label{sec:results}

\begin{figure*}
 \centering
 \includegraphics[width=0.95\textwidth]{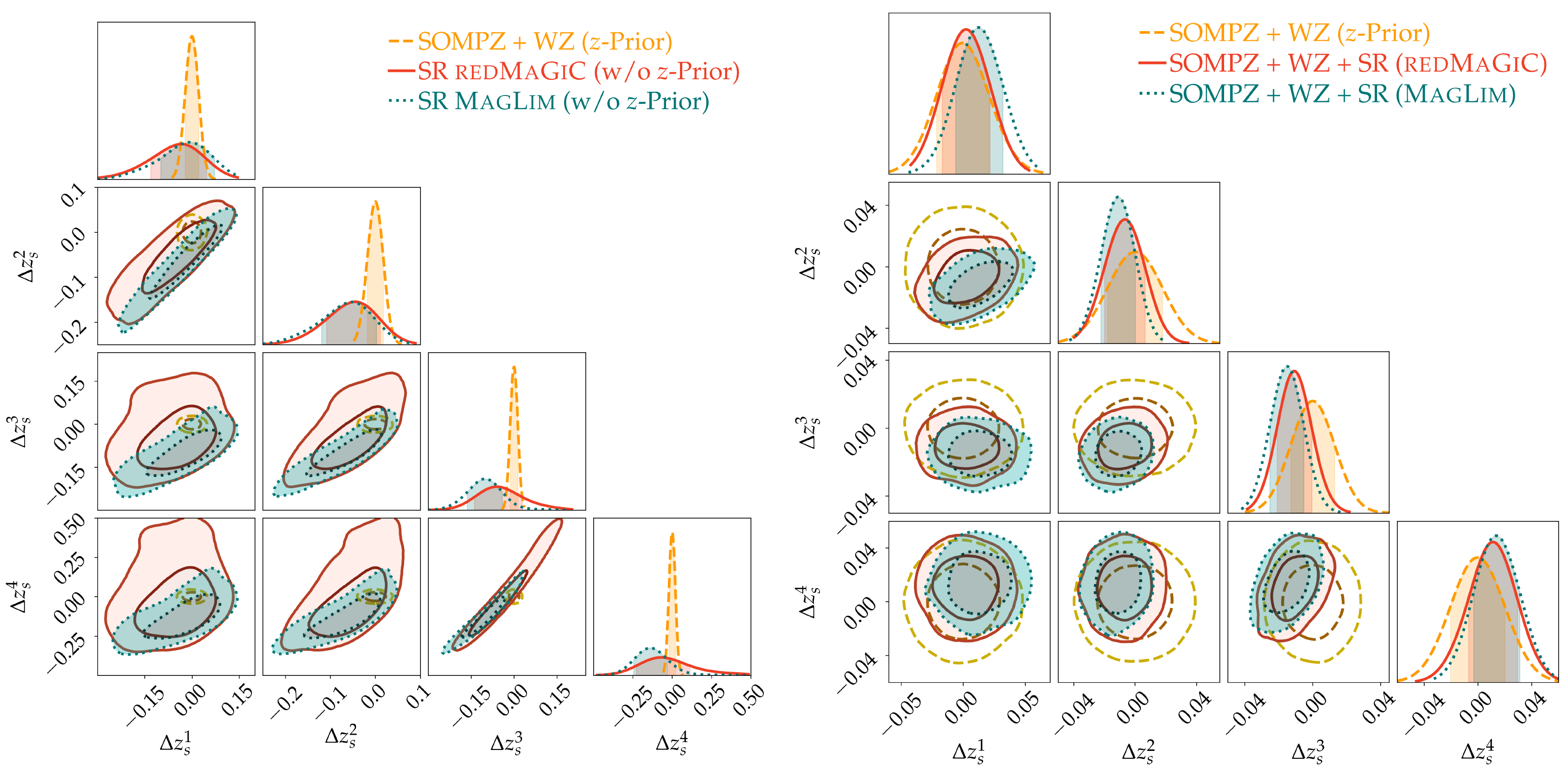}
 \caption{Mean source redshift constraints from a shear-ratio only chain (SR), with a flat uninformative prior, in comparison with the results from the combination of the alternative calibration methods of SOMPZ + WZ, and the final combined results of SOMPZ+WZ+SR on data using the \textsc{redMaGiC} sample.}
 \label{fig:data_deltaz}
 \end{figure*}

\begin{figure}
 \centering
 \includegraphics[width=0.45\textwidth]{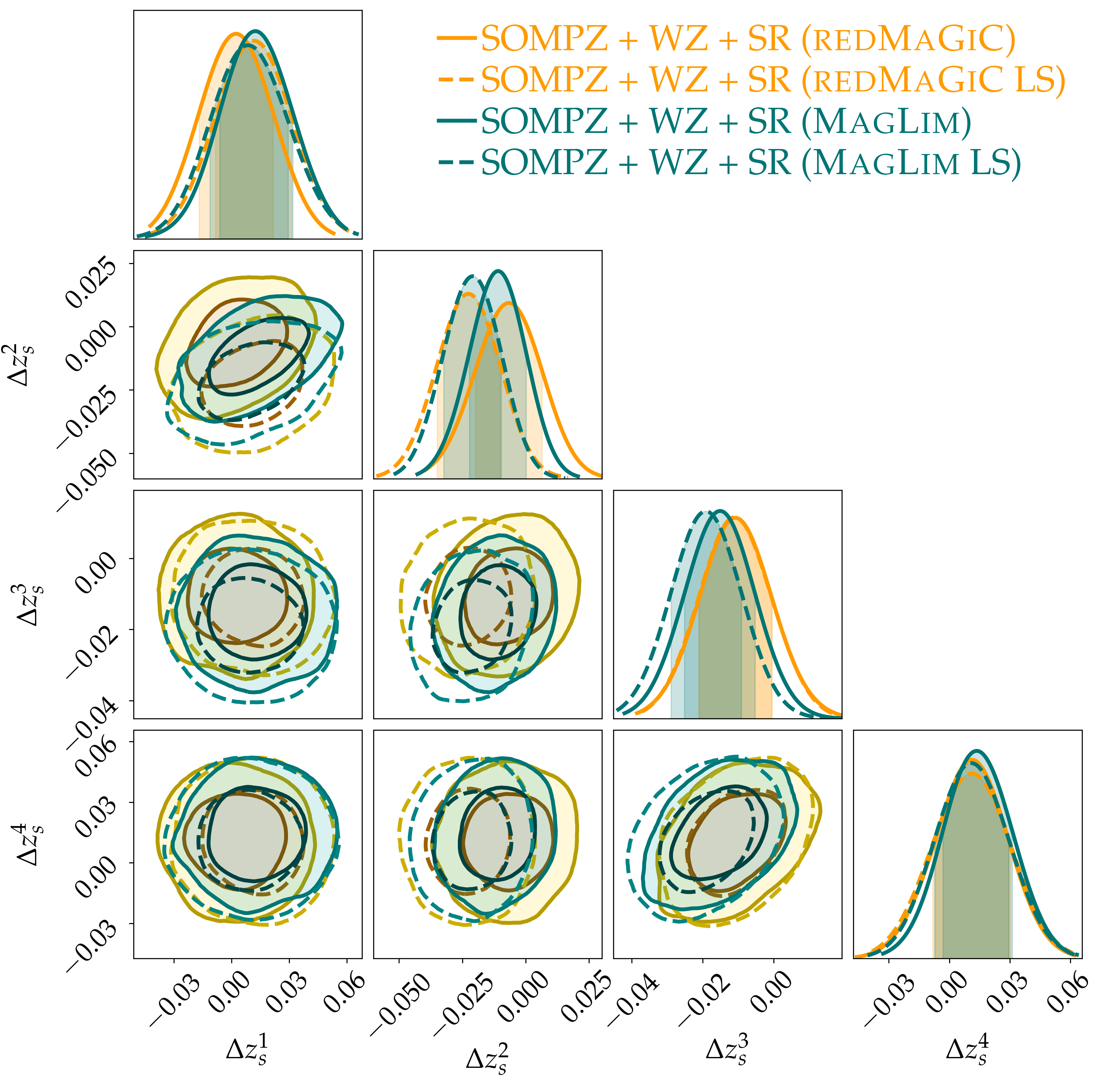}
 \caption{Mean source redshift constraints from different shear-ratio (SR) configurations, using the DES Y3 redshift prior (SOMPZ + WZ), comparing the fiducial small-scale constraints from those of the large-scale SR (LS), for the two independent lens galaxy samples, \textsc{redMaGiC} and \textsc{MagLim}.}
 \label{fig:data_deltaz_ls}
 \end{figure}
 
 \begin{figure}
 \centering
 \includegraphics[width=0.45\textwidth]{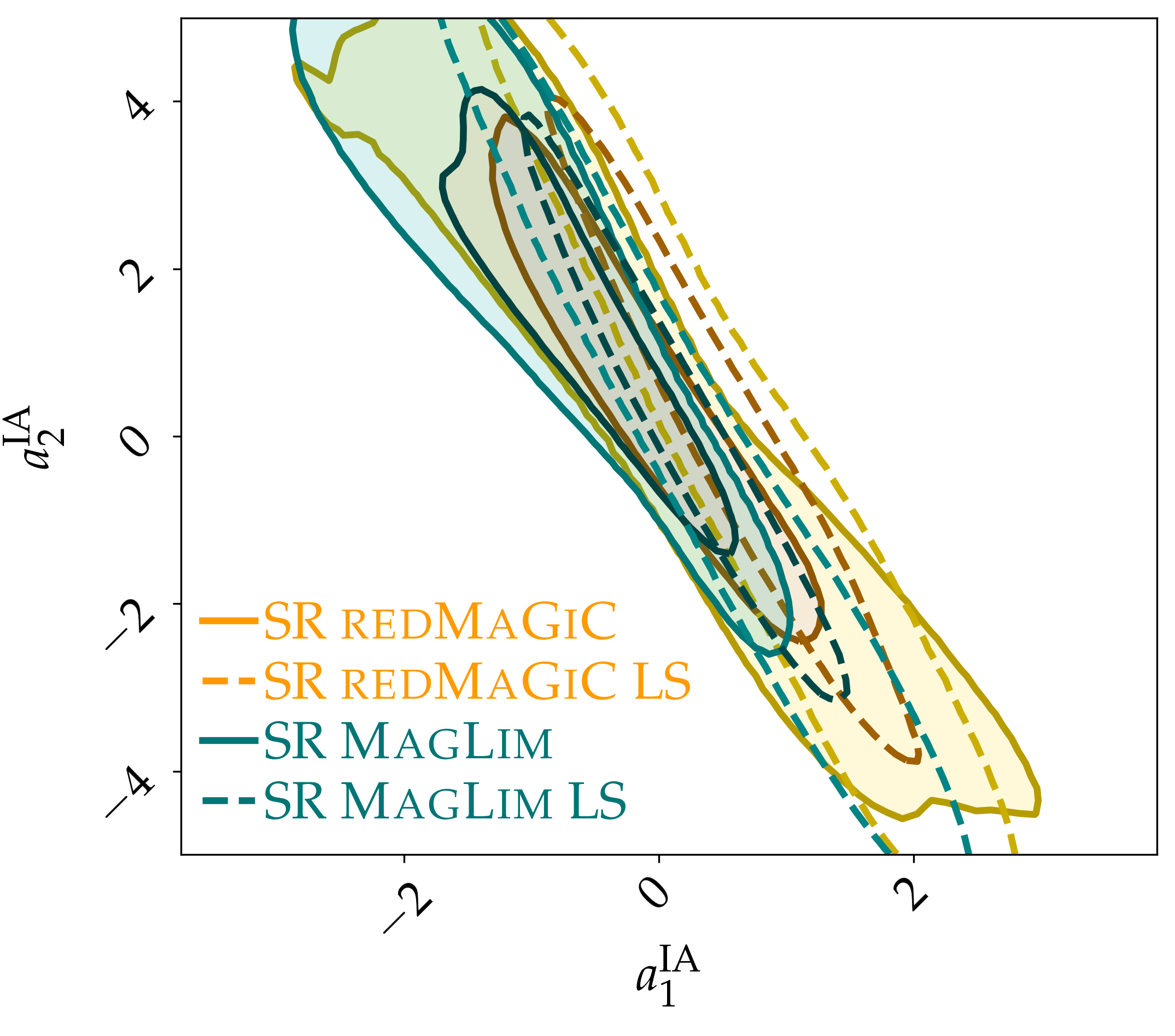}
 \caption{Data constraints on the two intrinsic alignment amplitude model parameters from different DES Y3 SR data configurations, comparing the fiducial small-scale constraints from those from the large-scale SR (LS), for the two independent lens galaxy samples, \textsc{redMaGiC} and \textsc{MagLim}. }
 \label{fig:ia_data}
 \end{figure}

In this section we will present and validate the constraints on model parameters derived from SR in the DES Y3 data sample. We compute SR for our two different lens samples, and for small and large scales. For a given set of ratios, $\{ r \}$, we use the following expression for computing the signal-to-noise:

\begin{equation}
    S/N = \sqrt{(\{ r \}) \: \mathbf{C}_{\{ r \}}^{-1} \: (\{ r \})^T - \mathrm{ndf}},
\end{equation}

where ndf is the number of degrees of freedom, which equals the number of ratios (9 in our case), and $\mathbf{C}$ is the covariance described in \S\ref{sec:cov}. Using the data ratios $\{ r \}_s$ (presented in Figure \ref{fig:ratios_sim_data}), we estimate, for the fiducial small-scale ratios, a combined $S/N \sim 84$ for the \textsc{MagLim} sample ($S/N \sim 60$ for the \textsc{redMaGiC} sample), and for large-scale ratios we estimate $S/N \sim 42$ for the \textsc{MagLim} sample ($S/N \sim 38$ for the \textsc{redMaGiC} sample).

We will broadly split the section in two parts: First, we will describe the model parameter constraints from SR alone, specifically by looking at their impact on source redshift and IA parameters, and study their robustness by using two different lens samples (\textsc{redMaGiC} and \textsc{MagLim}) and large-scale ratios for validation. Then, we will study the impact of SR in improving model parameter constraints when combined with other probes such as cosmic shear, galaxy clustering and galaxy-galaxy lensing in the DES Y3 data sample. It is worth pointing out that SR will also be used for correlations between DES data and CMB lensing, although these will not be discussed here (see \citealt{y3-cmblensing1,y3-cmblensing2} for the usage of SR in combination with CMB lensing). For the results in this Section, unless we specifically note that we free some of these priors, we use the DES Y3 priors on the parameters of our model, summarized in Table~\ref{tab:data_priors}.

\begin{figure*}
 \centering
 \includegraphics[width=0.95\textwidth]{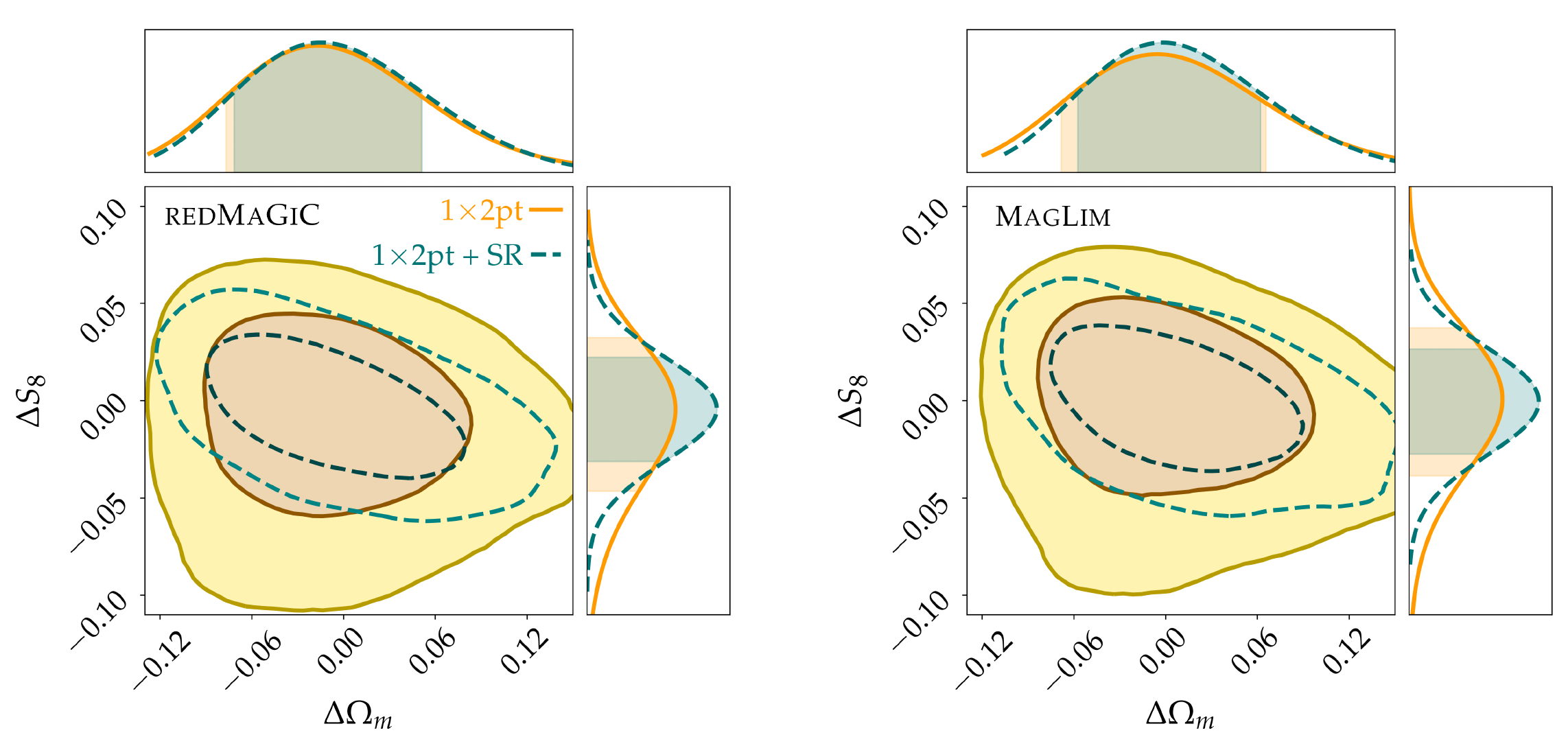}
 \caption{Differences in the DES Y3 data constraints on cosmological parameters $S_8$ and $\Omega_m$ with the addition of SR to the cosmic shear measurement ($1\times2$pt). The left panel shows the case with SR using \textsc{redMaGiC} lenses, while the right panel shows the results with SR using the \textsc{MagLim} lens sample. All the contours in the plot have been placed at the origin of the $\Delta \Omega_m$ -- $\Delta S_8$ plane, so that the plot shows only the impact of SR in the size of contours but does not include information on the central values of parameters or shifts between them. The impact of SR is significantly relevant for cosmic shear, with improvements in constraining $S_8$ of 31\% for \textsc{redMaGiC} SR and 25\% for \textsc{MagLim} SR. }
 \label{fig:cosmo_data}
 \end{figure*}

\begin{figure}
 \centering
 \includegraphics[width=0.45\textwidth]{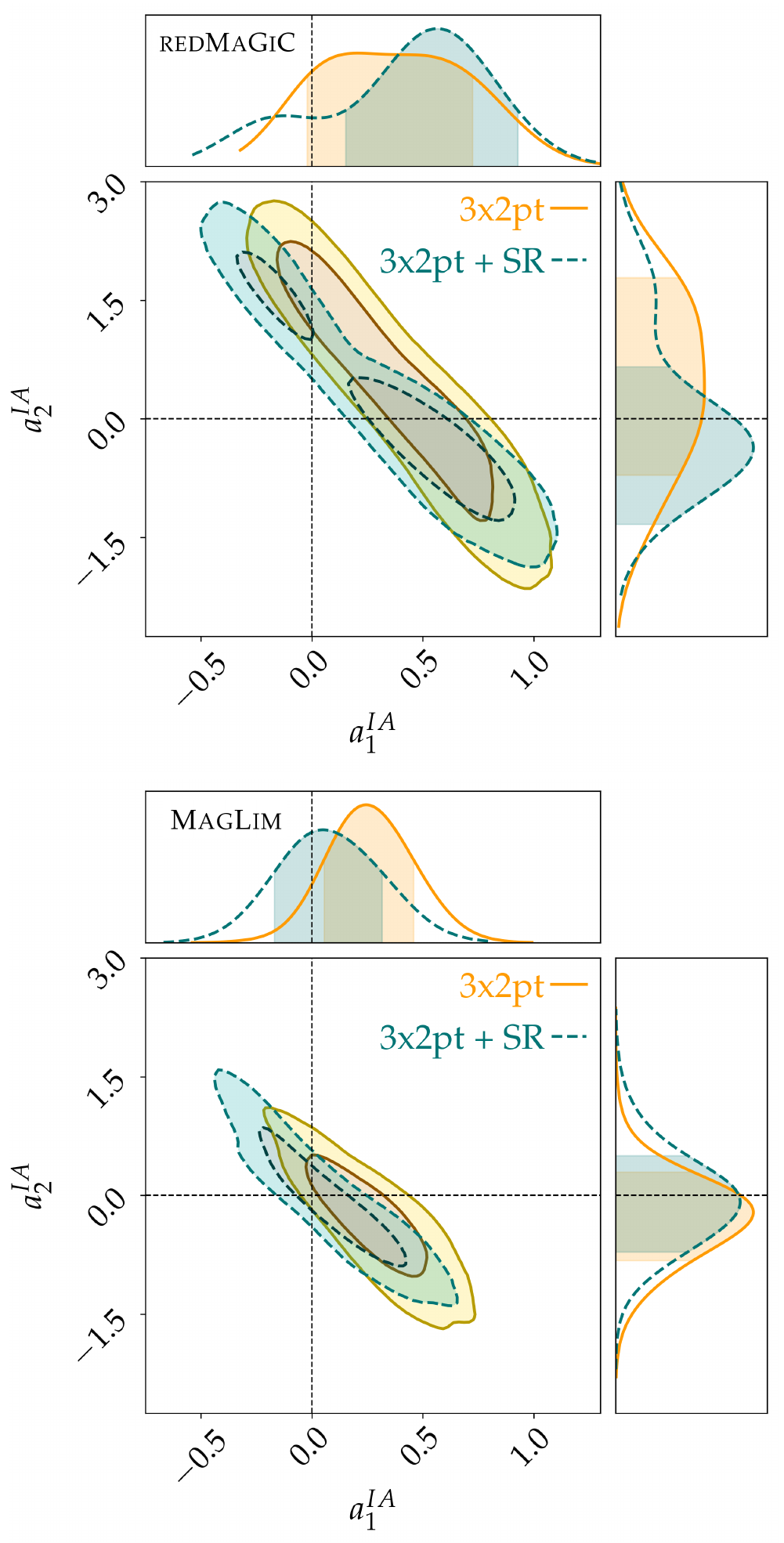}
 \caption{DES Y3 data constraints on the two intrinsic alignment amplitude model parameters from the full combination of probes (3$\times$2pt) with and without the addition of SR, for the \textsc{redMaGiC} and \textsc{MagLim} lens samples. The crossing of the dashed black lines shows the no IA case.}
 \label{fig:ia_3x2}
 \end{figure}

\subsection{DES Y3 SR-only constraints}\label{sec:results:SRonly}

\begin{table*}

\centering
\addtolength{\tabcolsep}{10pt} 
\begin{tabular}{l|c|c}
                     &         \textbf{Range}             &        \textbf{ Data Priors}             \\
\rule{0pt}{3ex}Source redshifts $\Delta z_s^j$                     &         [-0.1, 0.1]             &        $\mathcal{N}$(0, [0.018,0.015,0.011,0.017])              \\
Shear calibration $m^j$                     &         [-0.1, 0.1]             &       $\mathcal{N}$([-0.0063, -0.0198, -0.0241, -0.0369], [0.0091,0.0078,0.0076,0.0076])               \\
Lens redshifts \textsc{redMaGiC} $\Delta z_l^i$                     &         [-0.05, 0.05]             &      $\mathcal{N}$([0.006, 0.001, 0.004], [0.004,0.003,0.003])                \\
Lens redshifts \textsc{MagLim} $\Delta z_l^i$                     &         [-0.05, 0.05]             &      $\mathcal{N}$([-0.009, -0.035, -0.005], [0.007,0.011,0.006])                \\
Lens redshifts \textsc{MagLim} $\sigma_{z_l}^i$                     &         [0.1, 1.9]             &      $\mathcal{N}$([0.975, 1.306, 0.87], [0.062,0.093,0.054])                \\
Galaxy bias $b^i$                     &         [0.8, 3.0]             &     Uniform                \\
IA $a_1, a_2, \alpha_1, \alpha_2$                     &         [-5, 5]             &     Uniform                 \\
IA bias TA                     &         [0, 2]             &     Uniform                 
\end{tabular}
\addtolength{\tabcolsep}{-10pt} 
\caption{Allowed ranges and priors of the model parameters for the DES Y3 data chains run in Section \ref{sec:results}. Indices $i$ in the labels refer to the 3 lens redshift bins, and indices $j$ refer to the 4 source redshift bins, all defined in Section \ref{sec:model}.  } \label{tab:data_priors}
\end{table*}

Now we will present and discuss the model parameter constraints from SR in DES Y3. Because we will show and compare the constraints from various SR configurations, including ratios from two independent lens samples, we will also assess the robustness of these results. As we demonstrated in \S\ref{sec:model_validation}, SR provides constraints on model parameters corresponding to source redshifts and intrinsic alignments, so we will focus on those for this part.

Figure \ref{fig:data_deltaz} presents the SR constraints on the source redshift parameters of our model using a number of SR configurations. The left panel shows the SR constraints coming from the independent \textsc{redMaGiC} and \textsc{MagLim} galaxy samples, using flat, uninformative priors on the source redshift parameters and the priors described in Table~\ref{tab:data_priors} for the rest of the parameters. In that panel, for comparison, we also include the source redshift prior used in the DES Y3 analysis, which comes from a combination of photometric information (SOMPZ) and clustering redshifts (WZ), and which is presented in detail in \citet*{y3-sompz} and \citet*{y3-sourcewz} and shown here in Table~\ref{tab:data_priors}. At this point we can compute the tension between SR redshift constraints and the redshift prior, for the 4 source redshift bins combined, and we obtain a $0.97\sigma$ tension ($p$-value $>$ 0.33) for the \textsc{redMaGiC} SR, and $2.08\sigma$ ($p$-value $>$ 0.04) for the \textsc{MagLim} (numbers computed following \citealt*{Lemos2020}). Since these values are above our threshold for consistency ($p$-value $>$ 0.01), the SR constraints are in agreement with the prior and we can proceed to use the redshift prior for the SR likelihoods (see \citealt*{y3-sompz} for a review of the complete DES Y3 weak lensing source calibration, and \citealt{y3-cosmicshear1} for SR consistency checks in combination with cosmic shear). Regarding the mild tension between SR and the redshift prior for the \textsc{MagLim} sample, we refer to \citet{y3-3x2ptkp} for results demonstrating the consistency of the cosmological constraints with and without SR. 

The right panel in Figure \ref{fig:data_deltaz} shows the \textsc{redMaGiC} and \textsc{MagLim} SR constraints when using the DES Y3 redshift prior, so we can visualize the improvement that SR brings to the prior redshift constraints. Specifically, for \textsc{redMaGiC} SR, the constraints on the 4 source redshift $\Delta z$ parameters are improved by 11\%, 28\%, 25\% and 14\% with respect to the prior, and for \textsc{MagLim}, by 14\%, 38\%, 25\% and 17\%, respectively for the 4 redshift parameters (the percentage numbers quote the reduction in the width of parameter posteriors compared to the prior). Note that within the DES Y3 3$\times$2pt setup, we do not use the SR information in this way, i.e., by using the redshift prior that comes from the combination of SOMPZ + WZ + SR, but instead we add the shear-ratio likelihood to the 3$\times$2pt likelihood as written in Eq.~(\ref{eq:combined_likelihood}). In this way, the SR information is not only constraining redshifts but also the rest of the parameters of the model, especially the parameters modeling IA.

The agreement between the SR constraints coming from our two independent lens samples, \textsc{redMaGiC} and \textsc{MagLim}, demonstrates the robustness of SR source redshift constraints and provides excellent validation for the methods used in this work. In addition, in Figure \ref{fig:data_deltaz_ls} we show the large-scale SR constraints for both lens samples, compared to the small-scale, fiducial SR constraints. As discussed in Sections \ref{sec:measurement} and \ref{sec:model_validation}, large-scale SR provides independent validation of the small-scale SR constraints. Because of the larger angular scales used in their calculation, they are less sensitive to effects such as non-linear galaxy bias or the impact of baryons (although we have demonstrated that small-scale ratios are also not significantly impacted by these in Sec.~\ref{sec:baryons_and_nlbias}). At this point we can again compute the tension between fiducial and large-scale SR, and we obtain a $0.1\sigma$ tension for the \textsc{redMaGiC} case, and $0.3\sigma$ for \textsc{MagLim} (numbers computed following \citealt*{Lemos2020}). This agreement between the fiducial small-scale SR and the large-scale versions, again for two independent lens galaxy samples, provides additional evidence of the robustness of the results in this work. 

In addition to the source redshift parameters, the other parameters that are significantly constrained by SR are the amplitudes of the IA model, $a_1^{\mathrm{IA}}$ and $a_2^{\mathrm{IA}}$ (see \S \ref{sec:model} for a description). Importantly, such constraints have a strong impact in tightening cosmological constraints when combined with other probes, such as cosmic shear (see Figure \ref{fig:1x2sim} and \citealt{y3-cosmicshear1}). Figure \ref{fig:ia_data} shows the IA amplitude constraints from \textsc{redMaGiC} and \textsc{MagLim} SR, both using small-scales (fiducial) and using large-scale (LS) SR as validation. The agreement between these constraints demonstrates the robustness of the IA SR constraints, which play an important role when combined with cosmic shear and other 2pt functions.

\subsection{Impact of SR on $1,2,3 \times 2$pt in the DES Y3 cosmological analysis}

The SR methods described in this work are part of the fiducial DES Y3 cosmological analysis, and hence the SR measurements are used as an additional likelihood to the other 2pt functions. In this part we will describe the impact of adding the SR likelihood in constraining our cosmological model when combined with other probes such as cosmic shear, galaxy clustering and galaxy-galaxy lensing. We will do so by comparing the cosmological constraints with and without SR and then describing the gains in constraining power in them when SR is used. Please note that we will focus on the gains of the combination with SR, and we will not present or discuss the cosmological results or their implications. For such presentation and discussion, please see the cosmic shear results in two companion papers \citet*{y3-cosmicshear1,y3-cosmicshear2}, the results from galaxy clustering and galaxy-galaxy lensing in \citet*{y3-2x2ptbiasmodelling,y3-2x2ptaltlensresults, y3-2x2ptmagnification} and the combination of all probes in \citet*{y3-3x2ptkp}.

Figure \ref{fig:cosmo_data} shows the impact of SR in constraining cosmological parameters $\Omega_m$ and $S_8$ when combined with cosmic shear data ($1\times2$pt) in the DES Y3 data, for both the SR case with \textsc{redMaGic} and \textsc{MagLim} lens samples. The contours in the plot have all been placed at the origin of the $\Delta \Omega_m$ -- $\Delta S_8$ plane, so that the plot shows only the impact of SR in the size of contours but does not include information on the central values. The gain in constraining power from the addition of the SR likelihood in the data is in line with our findings on noiseless simulated data (\S\ref{sec:combination} and Figure \ref{fig:1x2sim}), pointing to the robustness of the simulated analysis in reproducing the DES Y3 data. As in the simulated case, SR is especially important in constraining cosmology from cosmic shear, where it improves the constraints on $S_8$ by 31\% for \textsc{redMaGiC} SR and 25\% for \textsc{MagLim} SR. As explored in Figure \ref{fig:1x2sim}, the improvement comes especially from the ability of SR to place constraints on IA, which then breaks important degeneracies with cosmology in cosmic shear. Given this role of SR as a key component of cosmic shear in constraining IA and cosmology, it is worth exploring the role played by SR in cosmic shear for different models of IA. In this paper we have assumed the fiducial IA model (TATT) for all tests. For a study showing how SR impacts the cosmic shear constraints using different IA models, see the DES Y3 cosmic shear companion papers \citet{y3-cosmicshear1} and \citet*{y3-cosmicshear2}.  In summary, we find that SR improves IA constraints from cosmic shear for all IA models. When using TATT (which is a five-parameter IA model) or NLA with redshift evolution (which is a three-parameter IA model), SR significantly helps constraining $S_8$ due to the breaking of degeneracies with IA. For the simplest NLA model without redshift evolution (which is a one-parameter IA model), SR significantly tightens the IA constraints from cosmic shear, but the impact on $S_8$ is reduced due to the milder degeneracies between IA and $S_8$ for that case. 

In the combination with the other 2pt functions in the data, galaxy clustering and galaxy-galaxy lensing, the improvement coming from SR is less pronounced, as expected from our simulated analysis, but nonetheless for the full combination of probes (3$\times$2pt) the addition of SR results in the DES Y3 data constraints on $S_8$ being tighter by 10\% for the \textsc{redMaGiC} case and 5\% for \textsc{MagLim} (see also \citealt{y3-3x2ptkp}). The SR improvement on the $3\times2$pt cases is slightly higher than what we found in the simulated case (\S\ref{sec:combination}), which may be due to the fact that the covariance used in the data is different from the simulated case, as it was re-computed at the best-fit cosmology after the $3\times2$pt unblinding (see \citealt{y3-3x2ptkp} for more details).

Besides the impact of SR in cosmological constraints, it is important to stress that SR does significantly impact parameter posteriors on source redshifts and intrinsic alignments in all cases, even in the cases where the improvements in cosmology are mild or negligible. In particular, for the full combination of probes ($3\times2$pt), the cases with SR present tighter posteriors on the second and third source redshift parameters (the ones SR constraints best) by around 14\%, for both \textsc{redMaGiC} and \textsc{MagLim}. In addition, SR does have an impact on the posteriors on IA for the full combination of probes, as can be seen in Figure \ref{fig:ia_3x2}. In that plot, one can see how the addition of SR pulls the IA constraints closer to the no IA case (marked in the plot as a cross of dashed lines), for both lens samples. This is consistent with Figure \ref{fig:ia_data}, where the SR data is shown to be consistent with the case of no IA, although in a degeneracy direction between IA parameters $a_1$ and $a_2$, and it demonstrates the impact of SR in the IA constraints even for the cases where such impact does not translate to a strong impact on cosmological constraints. For a discussion of IA in the context of the $3\times2$pt analysis, see \citet{y3-3x2ptkp}.

\section{Summary and conclusions}
\label{sec:conclusions}

The Dark Energy Survey Y3 3$\times$2pt cosmological analysis, much like other cosmological analyses of photometric galaxy surveys, relies on the combination of three measured 2pt correlation functions, namely galaxy clustering, galaxy-galaxy lensing and cosmic shear. The usage of these measurements to constrain cosmological models, however, is limited too large angular scales because of the uncertainties coming from modeling baryonic effects and galaxy bias. Consequently, a significant amount of information at smaller angular scales typically remains unused in these analyses.

In this work we have developed a method to use small-scale ratios of galaxy-galaxy lensing measurements to place constraints on parameters of our model, particularly those corresponding to source redshift calibration and intrinsic alignments. These ratios of galaxy-galaxy lensing measurements, evaluated around the same lens bins, are also known as lensing or shear ratios (SR). The SR have often been used in the past assuming they were a purely geometrical probe. In this work, instead, we use the full modeling of the galaxy-galaxy lensing measurements involved, including the corresponding integration  over  the  power  spectrum  and  the  contributions  from  intrinsic  alignments  and lens weak lensing magnification. Taking ratios of small-scale galaxy-galaxy lensing measurements sharing the same lens bins reduces their sensitivity to non-linearities in galaxy bias or baryonic effects, but retains crucial and independent information about redshift calibration and effects on intrinsic alignments, which we fully exploit with this approach.

We perform extensive testing of the small-scale shear ratio modeling by characterizing the impact of different effects, such as the inclusion of baryonic physics in the power spectrum, non-linear galaxy biasing, the effect of HOD modeling description and lens magnification. We test the shear ratio constraints on realistic $N$-body simulations  of  the  DES  data. We find that shear ratios as defined in this work are not significantly affected by any of those effects. We also use simulated data to study the constraining power of SR given the DES Y3 modeling choices and priors, and find it to be most sensitive to the calibration of source redshift distributions and to the amplitude of intrinsic alignments (IA) in our model. In particular, the sensitivity to IA makes SR very important when combined with other probes such as cosmic shear, and SR can significantly improve the constraints on cosmological parameters by breaking their degeneracies with IA. 

The shear ratios presented in this work are utilized as an additional contribution to the likelihood for cosmic shear and the full 3$\times$2pt in the fiducial DES Y3 cosmological analysis. The SR constraints have an important effect in improving the constraining power in the analysis. Assuming four source galaxy redshift bins, SR improves the constraints on the mean redshift parameters of those bins by up to more than 30\%. For the cosmic shear analysis, presented in detail in two companion papers \citet{y3-cosmicshear1} and \citet*{y3-cosmicshear2}, we find that SR improves the constraints on the amplitude of matter fluctuations $S_8$ by up to 31\%, due to the tightening of redshift posteriors but especially due to breaking degeneracies with intrinsic alignments (IA). For the full combination of probes in DES Y3 data, the so-called $3\times2$pt analysis \citep{y3-3x2ptkp}, SR improves the constraints on $S_8$ by up to 10\%. Even for the cases where the improvements in cosmology are mild, SR brings significant and independent information to the characterization of IA and source redshifts. In addition, when adding CMB lensing information to the DES Y3 analysis, \citet{y3-cmblensing1,y3-cmblensing2} find significant improvements with the addition of SR to the cross-correlation between shear and CMB lensing convergence maps, again due to constraints on intrinsic alignments.

One of the main advantages of SR is its weak sensitivity to modeling uncertainties at small scales, compared to the pure galaxy-galaxy lensing measurements. For that reason, for any choice of angular scales performed for galaxy-galaxy lensing, there will always be smaller angular scales that will be available for SR. These scales can be used to extract independent information. In addition, SR is more sensitive to the mean redshift rather than the width or shape of the full redshift distribution, complementing other methods (such as clustering cross-correlations) that are more sensitive to other moments of these distributions. Even more importantly, SR  provides redshift calibration even when the redshift distributions do not overlap with spectroscopic samples used for clustering cross-correlations, providing valuable independent information.

For these reasons, we conclude that SR can become a standard addition to cosmological analyses from imaging surveys using cosmic shear and 3$\times$2-like data. Furthermore, if redshift and intrinsic alignment modeling does not improve as quickly as the increased quality and quantity of data, then SR may become even more important for cosmological inference than it has been in DES Y3. This scenario seems likely given that source redshift priors did not improve significantly between Y1 and Y3, and the model of intrinsic alignments moved from 3 to 5 parameters from Y1 to Y3, thus becoming more complicated. Therefore, it seems plausible that SR will become an important tool to characterize these two uncertainties in our model, and hence become even more relevant at improving the cosmological constraints in future analyses. 

\section*{Acknowledgments}

CS is supported by grants AST-1615555 from the U.S. National Science Foundation, and DE-SC0007901 from the U.S. Department of Energy (DOE). JP is supported by DOE grant DE-SC0021429. Funding for the DES Projects has been provided by the U.S. Department of Energy, the U.S. National Science Foundation, the Ministry of Science and Education of Spain, 
the Science and Technology Facilities Council of the United Kingdom, the Higher Education Funding Council for England, the National Center for Supercomputing 
Applications at the University of Illinois at Urbana-Champaign, the Kavli Institute of Cosmological Physics at the University of Chicago, 
the Center for Cosmology and Astro-Particle Physics at the Ohio State University,
the Mitchell Institute for Fundamental Physics and Astronomy at Texas A\&M University, Financiadora de Estudos e Projetos, 
Funda{\c c}{\~a}o Carlos Chagas Filho de Amparo {\`a} Pesquisa do Estado do Rio de Janeiro, Conselho Nacional de Desenvolvimento Cient{\'i}fico e Tecnol{\'o}gico and 
the Minist{\'e}rio da Ci{\^e}ncia, Tecnologia e Inova{\c c}{\~a}o, the Deutsche Forschungsgemeinschaft and the Collaborating Institutions in the Dark Energy Survey. 

The Collaborating Institutions are Argonne National Laboratory, the University of California at Santa Cruz, the University of Cambridge, Centro de Investigaciones Energ{\'e}ticas, 
Medioambientales y Tecnol{\'o}gicas-Madrid, the University of Chicago, University College London, the DES-Brazil Consortium, the University of Edinburgh, 
the Eidgen{\"o}ssische Technische Hochschule (ETH) Z{\"u}rich, 
Fermi National Accelerator Laboratory, the University of Illinois at Urbana-Champaign, the Institut de Ci{\`e}ncies de l'Espai (IEEC/CSIC), 
the Institut de F{\'i}sica d'Altes Energies, Lawrence Berkeley National Laboratory, the Ludwig-Maximilians Universit{\"a}t M{\"u}nchen and the associated Excellence Cluster Universe, 
the University of Michigan, the National Optical Astronomy Observatory, the University of Nottingham, The Ohio State University, the University of Pennsylvania, the University of Portsmouth, 
SLAC National Accelerator Laboratory, Stanford University, the University of Sussex, Texas A\&M University, and the OzDES Membership Consortium.

The DES data management system is supported by the National Science Foundation under Grant Numbers AST-1138766 and AST-1536171.
The DES participants from Spanish institutions are partially supported by MINECO under grants AYA2015-71825, ESP2015-88861, FPA2015-68048, SEV-2012-0234, SEV-2016-0597, and MDM-2015-0509, 
some of which include ERDF funds from the European Union. IFAE is partially funded by the CERCA program of the Generalitat de Catalunya.
Research leading to these results has received funding from the European Research
Council under the European Union's Seventh Framework Program (FP7/2007-2013) including ERC grant agreements 240672, 291329, and 306478.
We  acknowledge support from the Australian Research Council Centre of Excellence for All-sky Astrophysics (CAASTRO), through project number CE110001020.

This manuscript has been authored by Fermi Research Alliance, LLC under Contract No. DE-AC02-07CH11359 with the U.S. Department of Energy, Office of Science, Office of High Energy Physics. The United States Government retains and the publisher, by accepting the article for publication, acknowledges that the United States Government retains a non-exclusive, paid-up, irrevocable, world-wide license to publish or reproduce the published form of this manuscript, or allow others to do so, for United States Government purposes.

Based in part on observations at Cerro Tololo Inter-American Observatory, 
National Optical Astronomy Observatory, which is operated by the Association of 
Universities for Research in Astronomy (AURA) under a cooperative agreement with the National 
Science Foundation.

\bibliography{library}
\bibliographystyle{mnras_2author}

\appendix

\section{Shear ratio covariance} \label{sec:appendix_covariance}

\begin{figure}
 \centering
 \includegraphics[width=0.45\textwidth]{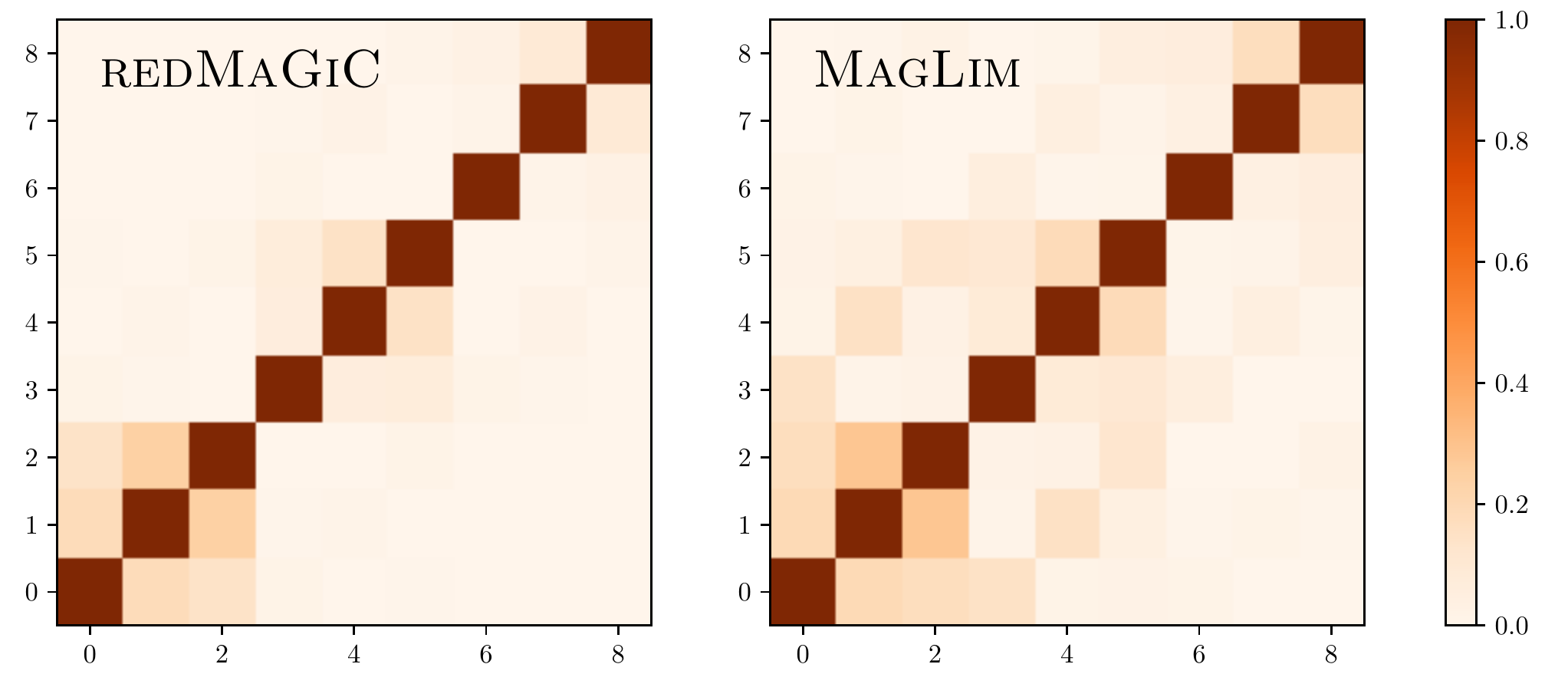}
 \caption{Correlation matrix for the lensing ratios, on the left panel using the \textsc{redMaGiC} lens sample and on the right panel using the \textsc{MagLim} sample. }
 \label{fig:ratios_cov}
 \end{figure}

Figure \ref{fig:ratios_cov} shows the covariance of the measured ratios, for the \textsc{redMaGiC} and \textsc{MagLim} ratios, following the procedure described in Section \ref{sec:cov}.

\section{Constraints on full IA model}
\label{sec:app_ia}
\begin{figure*}
 \centering
 \includegraphics[width=0.8\textwidth]{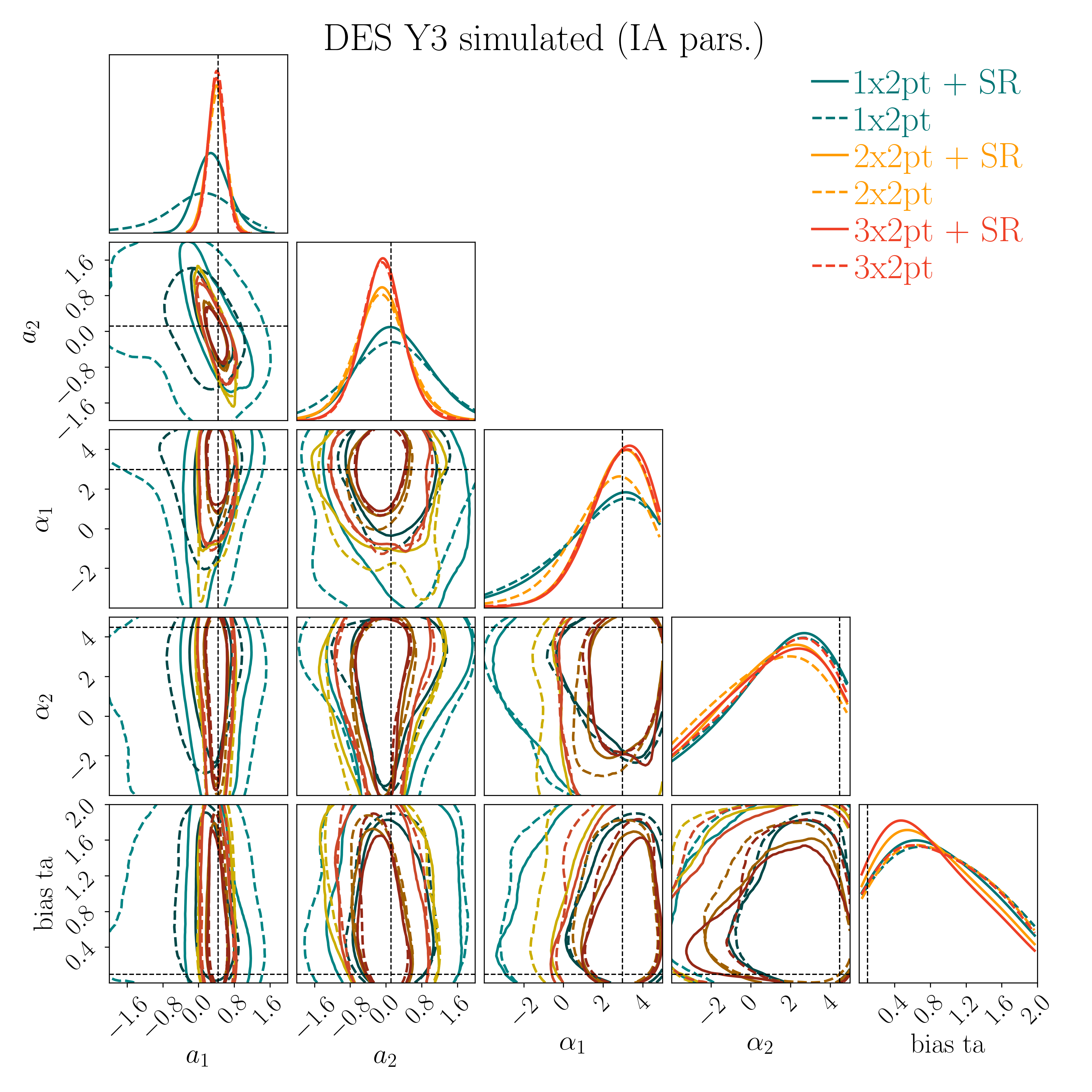}
 \caption{Constraints on the five parameters of the IA model described in \S\ref{sec:modeling} given the combination of SR and the other 2pt functions, using simulated DES Y3 data. }
 \label{fig:app_ia}
 \end{figure*}

Throughout the paper we have extensively discussed the SR constraints on two IA parameters of our model, $a_1$ and $a_2$, because they are the two parameters SR constrains best. In Figure \ref{fig:app_ia}, for completeness, we show the impact of SR in constraining all five parameters of the IA model (described in \S\ref{sec:modeling}) when combined with other 2pt functions. We can recognize the strong impact of SR for cosmic shear ($1\times2$pt), especially in the $a_1$ -- $a_2$ plane, but for the other parameters we can see that the impact of SR is not very significant. 

\section{HOD model of galaxy-galaxy lensing}\label{app:HOD}

In this appendix we describe the prediction for galaxy-galaxy lensing using a halo model framework. As described in \S~\ref{sec:hod}, we measure the mean HOD of all the galaxies in each tomograhic bin as well as in a fine binning of $\delta z \sim 0.02$ to capture the effects of evolution of HOD within the redshift bin. For each tomographic sub-bin, we measure both the number of central galaxies ($N_{\rm cen}$) and number of satellite galaxies ($N_{\rm sat}$) that we use in the modeling below. In order to estimate the galaxy-galaxy lensing signal from these measurements, we first predict the 1-halo and 2-halo angular power spectrum between the galaxy position and the convergence fields which can be written as follows. The 1-halo contribution is given by:
\begin{multline}\label{eq:Cl1h}
C_{\rm g\kappa,1h}^{ij}(\ell) = \int_{z_{\rm{min}}}^{z_{\rm{max}}} dz \frac{dV}{dz d\Omega} \int_{M_{\rm{min}}}^{M_{\rm{max}}} dM \frac{dn}{dM} \\ \bar{u}_{\rm g}^{i}(\ell,M,z) \ \bar{u}_{\kappa}^{j}(\ell,M,z),
\end{multline}
where $dV$ is the cosmological volume element, $dn/dM$ is the halo mass function, and $\bar{u}_{\rm g}^{i}$ and $\bar{u}_{\kappa}^{j}$ are the multipole-space profiles of observables galaxy and convergence fields for tomgraphic bins $i$ and $j$ respectively. We use the \citet{Crocce_2010} fitting function for the halo mass function, $dn/dM$ throughout. 

The two-halo term is given by:
\begin{equation}\label{eq:Cl2h}
C_{\rm g\kappa,2h}^{ij}(\ell) = \int_{z_{\rm{min}}}^{z_{\rm{max}}} dz \frac{dV}{dz d\Omega} b_{\rm g}^{i}(\ell, z) \ b_{\kappa}^{j}(\ell, z) \ P_{\rm{lin}}((\ell + 1/2)/\chi,z),
\end{equation}
where $b_{\rm g}^{i}$ and $b_{\kappa}^{j}$ are effective linear bias parameters describing the clustering of galaxy and convergence field respectively, $P_{\rm lin}(k,z)$ is the linear matter power spectrum and $\chi$ is the comoving distance corresponding to the redshift $z$. 

The multipole space profile of the dark matter distribution is given by:
\begin{equation}\label{eq:ukl}
    \bar{u}_{\kappa}^{j}(\ell,M,z) = \frac{W^j_\kappa(z)}{\chi^2}  u_{\rm{m}}(k,M),
\end{equation}
where $k = (\ell + 1/2) / \chi$, and $W^j_\kappa(z)$ is the lensing efficiency of source galaxies corresponding to redshift bin $j$ as defined in Eq.~\ref{eq:lensing_efficiency}. Here we approximate $u_{\rm{m}}(k,M)$ with a Navarro-Frenk-White (NFW) profile and use the concentration relation from \citet{Bullock_2001} to predict it.

The multipole space profile of the galaxy distribution is related to $\bar{u}_{\kappa}^{j}(\ell,M,z)$ and is given by:
\begin{multline}\label{eq:ugl}
    \bar{u}_{\rm g}^{i}(\ell,M,z) = \frac{W^i_g(z)}{\chi^2} \frac{1}{\langle n_g(z) \rangle} \times \\  (N_{\rm{cen}}(M,z) +  N_{\rm{sat}}(M,z) u_{\rm{sat}}(k,M)) , 
\end{multline}
where $W^i_g = (dn^i_g/dz)(dz/d\chi)$ with $(dn^i_g/dz)$ the normalized redshift distribution of the galaxies corresponding to redshift bin $i$, $N_{\rm cen}$ and $N_{\rm sat}$ are the central and satellite galaxy numbers.  We assume that the spatial distribution of satellite galaxies, $u_{\rm sat}$, can be approximated by the same NFW profile as matter, $u_{\rm{sat}} = u_{\rm{m}}$.

For the 2-halo term, the effective linear bias of the dark matter halos can be written as:
\begin{equation}\label{eq:bkl}
b^{\kappa,j}(\ell,z)=\frac{W^j_\kappa(z)}{\chi^2} \ \int dM \frac{dn}{dM} b_{\rm{lin}}(M,z) u_{\rm{m}}(k,M_{\rm{vir}})\ .
\end{equation}
We approximate the linear bias of halos $b_{\rm{lin}}$ with the \citet{Bhattacharya_2011} fitting function. 

The mean number of galaxies, $\langle n_g(z) \rangle$, entering into Eq.~\ref{eq:ugl}, is then given by:
\begin{equation}\label{eq:ng}
\langle n_g(z) \rangle = \int_{M_{\rm{{min}}}}^{M_{\rm{{max}}}} dM \frac{dn}{dM} (M) (N_{\rm{cen}}(M,z) + N_{\rm{sat}}(M,z)),
\end{equation}
where $M_{\rm{{min}}}$ and $M_{\rm{{max}}}$ correspond to the boundaries of a particular mass bin.  Similarly, the effective large scale bias of the galaxies is given by:
\begin{multline}\label{eq:bgl}
    b_{g}^{i}(\ell,z) = \frac{W^i_g(z)}{\chi^2} \frac{1}{\langle n_g(z) \rangle} \int_{M_{\rm{{min}}}}^{M_{\rm{{max}}}} dM \frac{dn}{dM} \\
     (N_{\rm{cen}}(M,z) + N_{\rm{sat}}(M,z))u_{\rm{sat}}(k,M,z))b_{\rm{lin}}(M,z).
\end{multline}

Finally, the galaxy-galaxy lensing signal in real space is given by:
\begin{equation}
    \gamma^{ij}_t (\theta) = \int \frac{d \ell \ell}{2\pi} J_2(\ell \theta) (C_{\rm g\kappa,1h}^{ij}(\ell)  + C_{\rm g\kappa,2h}^{ij}(\ell)),
\end{equation}
where $J_2$ is the second order Bessel function of the first kind. 

We use this framework to predict the galaxy-galaxy lensing signal and hence the corresponding shear ratios between different redshift bins. To that end, we use the measured $N_{\rm{cen}}$ and $N_{\rm{sat}}$ from the DES galaxy mock catalogs as described in \citet{Crocce2015a, MacCrann_2019}. In  Fig.~\ref{fig:hod}, we show the impact of small-scale physics parameters parameterized by this HOD framework on the inferred shear ratios. We compare the case of assuming a constant HOD within a redshift bin in top panel and including the evolution of the $N_{\rm{cen}}$ and $N_{\rm{sat}}$ parameters within each redshift bins in bottom panel. We find a small impact of the small scale physics, particularly on the large scale shear ratios as quantified in the $\Delta \chi^2$ mentioned in the legend of the plot.

\label{lastpage}
\end{document}